\documentclass[12pt,draftcls,onecolumn]{IEEEtran} 
\usepackage{enumerate}
\usepackage{times} 
\usepackage{latexsym}
\usepackage{xspace}
\usepackage{amssymb,amsmath,amsfonts,color}
\usepackage{mathrsfs}
\usepackage{cite}
\usepackage{slashbox}
\usepackage{subfigure}
\usepackage{epsfig} 
\usepackage{psfrag}

\usepackage{url}

\newtheorem{corollary}{Corollary}
\newtheorem{definition}{Definition}

\newtheorem{lemma}{Lemma}

\newtheorem{proposition}{Proposition}

\newtheorem{theorem}{Theorem}

\title{Consensus of Multi-Agent Networks in the Presence of Adversaries Using Only Local Information}
\author{Heath LeBlanc, Haotian Zhang, Shreyas Sundaram, and Xenofon Koutsoukos
\thanks{Heath LeBlanc and Xenofon Koutsoukos are with the Department of Electrical Engineering and Computer Science at Vanderbilt University. Email for corresponding author: {\tt heath.j.leblanc@vanderbilt.edu}.}%
\thanks{Haotian Zhang and Shreyas Sundaram are with the Department of Electrical and Computer Engineering at the University of Waterloo. Email: {\tt ssundara@uwaterloo.ca}.}%
\thanks{This report contains the proofs of the results presented at the First Conference on High-Confidence Networked Systems (HiCoNS 2012) \cite{LeBlanc2012_ConsOfMANsInPresOfAdvUsingOnlyLocalInfo}.}
}

\begin{document}
\maketitle
\thispagestyle{empty}
\pagestyle{empty}

\begin{abstract}
This paper addresses the problem of resilient consensus in the presence of misbehaving nodes. Although it is typical to assume knowledge of at least some nonlocal information when studying secure and fault-tolerant consensus algorithms, this assumption is not suitable for large-scale dynamic networks. To remedy this, we emphasize the use of local strategies to deal with resilience to security breaches. We study a consensus protocol that uses only local information and we consider worst-case security breaches, where the compromised nodes have full knowledge of the network and the intentions of the other nodes. We provide necessary and sufficient conditions for the normal nodes to reach consensus despite the influence of the malicious nodes under different threat assumptions. These conditions are stated in terms of a novel graph-theoretic property referred to as \textit{network robustness}. 
\end{abstract}


\section{\uppercase{Introduction}}
\label{S:introduction}
The engineering community has experienced a paradigm shift from centralized to distributed system design, propelled by advances in networking and low-cost, high performance embedded systems.  In particular, this has led to significant interest in the design and analysis of \emph{multi-agent networks}. A multi-agent network consists of a set of individuals called \emph{agents}, or \emph{nodes}, equipped with some means of sensing or communicating along with computational resources and possibly actuation. Through a medium, which is referred to as the \emph{network}, the agents share information in order to achieve specific \emph{group objectives}. Some examples of group objectives include consensus~\cite{OlfatiSaber07_ConsensusAndCoopInNetMultiagentSys,Ren07_InfoConsInMultiVehCoopCtrl}, synchronization~\cite{Scardovi_SyncNetsIdentLinSys_Automatica09, Chopra06_PBCtrlOfMultiAgSys}, surveillance~\cite{Casbeer06_CoopForestFireSurvUsingTeamUAVs}, and formation control~\cite{Fax04_InfoFlowCoopCtrlVehForm}. In order for the group objectives to be achieved, \emph{distributed algorithms} are used to coordinate the behavior of the agents.

There are several advantages to using multiple agents over a single one. First, the objective may be complex and challenging, or possibly even infeasible for a single agent to achieve. Second, employing many agents can provide robustness in the case of failures or faults. Third, networked multi-agent systems are flexible and can support reconfigurability. Finally, there are performance advantages that can be leveraged from multiple agents. For example, in surveillance and monitoring applications, a multi-agent network provides redundancy and increased fidelity of information \cite{Casbeer06_CoopForestFireSurvUsingTeamUAVs,Kingston08_DecPerSurvUsingTeamOfUAVs}.

Along with the advantages come certain challenges. Perhaps the most fundamental challenge in the design of networked multi-agent systems is the restriction that the coordination algorithms use only \emph{local information}, i.e., information obtained by the individual agent through sensor measurements, calculations, or communication with neighbors in the network. In this manner, the feedback control laws must be \emph{distributed}.

A second challenge lies in the fact that not only is each agent typically a dynamical system, but the network itself is dynamic. This challenge arises because the agents may be mobile and the environment may be changing, thus giving rise to dynamic (or switching) networks. Since the distributed algorithms depend directly on the network, this additional source of dynamics can affect the stability and performance of the networked system.

An especially important challenge is that multi-agent networks, like all large-scale distributed systems, have many entry points for malicious attacks or intrusions. 
For the success of the group objective, it is important that the cooperative control algorithms are designed in such a way that they can withstand the compromise of a subset of the nodes and \emph{still guarantee some notion of correct behavior at a minimum level of performance}. We refer to such a multi-agent network as being \emph{resilient} to adversaries. Given the growing threat of malicious attacks in large-scale cyber-physical systems, this is an important and challenging problem~\cite{CardenasAminSastry08_ResChall4SecurityofCtrlSys}.

One of the most fundamental group objectives is to reach consensus on a quantity of interest. This concept is deeply intuitive, yet imprecise. Hence, there are several variations on how consensus problems are defined. At one extreme, consensus may be \emph{unconstrained}, and there is no restriction on the agreement quantity. In other cases, consensus may be \emph{partially constrained} by some rule or prescribed to lie in a set of possible agreement values which are in some way reasonable to the problem at hand. At the other extreme, consensus may be \emph{function constrained}, or $\chi$-constrained, in which case the consensus value must satisfy a particular function of the initial values of the nodes~\cite{Cortes_DistAlgs4ReachConsOnGenFcns08, Sundaram08_DistFuncCalcConsUsingLinItStrat}. In all of these cases, it is important that consensus algorithms be \emph{resilient} to various forms of uncertainty, whether the source of uncertainty is caused by implementation effects, faults, or security breaches. 

The problem of reaching consensus resiliently in the presence of misbehaving nodes has been studied in distributed computing~\cite{Lamport82_ByzGeneralProb,lynch96:dist_algs}, communication networks~\cite{Hromkovic05_DissOfInfoInCommNets}, and mobile robotics~\cite{Agmon06_FaultTolGathAlgs4AutMobRobots, Defago06_FTSSMobileRobGather, Bouzid2010_OptByzResConvInUniDimRobNets}. Among other things, it has been shown that given $F$ Byzantine or malicious nodes, there exists a strategy for the misbehaving nodes to disrupt consensus if the network connectivity\footnote{The network connectivity is defined as the smaller of the two following values: (i) the size of a minimal vertex cut and (ii) $n-1$, where $n$ is the number of nodes in the network.} is $2F$ or less.  Conversely, if the network connectivity is at least $2F+1$, then there exists strategies for the \emph{normal} nodes to use that ensure consensus is reached~\cite{lynch96:dist_algs, Sundaram2011_DistFunctCalcViaLinItStratInPresOfMalAgs, Pasqualetti2011_ConsCompInUnrelNets }. However, these methods either require that normal nodes have at least some nonlocal information or assume that the network is \emph{complete}, i.e., all-to-all communication or sensing~\cite{Lamport82_ByzGeneralProb, LeBlanc11_ConsNetMASWithAdv, Agmon06_FaultTolGathAlgs4AutMobRobots, Defago06_FTSSMobileRobGather, Bouzid2010_OptByzResConvInUniDimRobNets}. Moreover, these algorithms tend to be computationally expensive. Therefore, there is a need for resilient consensus algorithms that are \emph{low complexity} and \emph{operate using only local information}.

Typically, an upper bound on the number of faults or threats in the network is assumed, i.e., at most $F$ out of $n$ nodes fail or are compromised. 
We refer to this \emph{threat assumption}, or \emph{scope of threat}, as the \emph{$F$-total model}. In cases where it is preferable to make \emph{no global assumptions}, we are interested in other threat assumptions that are strictly local. For example, whenever each node only assumes that at most $F$ nodes in its \emph{neighborhood} are compromised (but there is no other bound on the total number of compromised nodes), the scope of threat is \emph{$F$-local}. 

In addition to the \emph{number} of misbehaving nodes, one can consider various \emph{threat models} for the misbehaving nodes; examples include \emph{non-colluding}~\cite{Pasqualetti2011_ConsCompInUnrelNets}, \emph{malicious}~\cite{Pasqualetti2011_ConsCompInUnrelNets, Sundaram2011_DistFunctCalcViaLinItStratInPresOfMalAgs, LeBlanc11_ConsNetMASWithAdv}, or \emph{Byzantine}~\cite{Lamport82_ByzGeneralProb, Agmon06_FaultTolGathAlgs4AutMobRobots, LeBlanc_LowCompResConsAdv_HSCC12, Vaidya2012_ItAppByzConsInArbDigraphs} nodes. Non-colluding nodes are unaware of the network topology, which other nodes are misbehaving, or the states of non-neighboring nodes.  On the other hand, malicious nodes have full knowledge of the networked system and therefore, worst case behavior must be assumed. The only difference between malicious and Byzantine nodes lies in their capacity for deceit. Malicious nodes are unable to convey different information to different neighbors in the network, whereas Byzantine nodes can.

Recently, we have studied resilient algorithms in the presence of misbehaving nodes. In \cite{LeBlanc11_ConsNetMASWithAdv}, we introduce the Adversarial Robust Consensus Protocol (ARC-P) for consensus in the presence of malicious agents under the $F$-total model in continuous-time complete networks, with the agents also modeled in continuous time. The results of \cite{LeBlanc11_ConsNetMASWithAdv} are extended to both malicious and Byzantine threat models in networks with constrained information flow and dynamic network topology in \cite{LeBlanc_LowCompResConsAdv_HSCC12}. In \cite{Zhang2012_RobInfoDiffAlgs2LocBdAdv}, we study general distributed algorithms with $F$-local malicious adversaries, encompassing ARC-P. In \cite{LeBlanc_LowCompResConsAdv_HSCC12, Zhang2012_RobInfoDiffAlgs2LocBdAdv}, we show that  traditional graph theoretic properties such as connectivity and degree, which have played a vital role in characterizing the resilience of distributed algorithms (see \cite{lynch96:dist_algs, Sundaram2011_DistFunctCalcViaLinItStratInPresOfMalAgs}), are no longer adequate when the agents make purely local decisions (i.e., without knowing nonlocal aspects of the network topology). Instead, in~\cite{Zhang2012_RobInfoDiffAlgs2LocBdAdv} we introduce a novel topological property, referred to as {\it network robustness}, and show that this concept is highly effective at characterizing the ability of purely local algorithms to succeed.
Separate sufficient and necessary conditions are provided in \cite{Zhang2012_RobInfoDiffAlgs2LocBdAdv} for ARC-P to achieve resilient consensus in discrete time, and it is shown that the preferential attachment mechanism for generating complex networks produces robust graphs.

In this paper, we continue our study of resilient consensus in the presence of malicious nodes while using only local information. We are interested in partially constrained, asymptotic consensus in dynamic networks. To allow for multiple interpretations of the results, we formulate the problem in a setting common to discrete and continuous time for node dynamics and time-invariant or time-varying network topologies. We extend the Adversarial Robust Consensus Protocol (ARC-P) introduced in \cite{LeBlanc11_ConsNetMASWithAdv} to weighted networks. We then describe robust network topologies that are rich enough to enable resilience to malicious nodes, but are not too restrictive in terms of communication cost (i.e., number of communication links); in particular, we generalize the robustness property of \cite{Zhang2012_RobInfoDiffAlgs2LocBdAdv}. Given these topological properties, we fully characterize the consensus behavior of the normal nodes using ARC-P under the $F$-total model of malicious nodes, and provide, for the first time, a necessary and sufficient condition for the algorithm to succeed. Additionally, for the $F$-local threat model, we provide improved separate necessary and sufficient conditions for asymptotic agreement of the normal nodes in the presence of malicious nodes.

The rest of the paper is organized as follows. Section~\ref{S:Model_Problem} introduces the problem in a framework common to discrete and continuous time. Section~\ref{S:Algorithm} presents ARC-P in the unified framework. Section~\ref{S:Topologies} motivates the need for robust network topologies and introduces the formal definitions.  The main results are given in Section~\ref{S:Results}. A simulation example is presented in Section~\ref{S:Simulations}. Finally, some discussion is given in Section~\ref{S:Discussion}.



\section{\uppercase{Problem Formulation}}
\label{S:Model_Problem}
Consider a time-varying network modeled by the (finite, simple) \emph{directed graph}, or \emph{digraph},  $\mathcal{D}[t]=\{\mathcal{V},\mathcal{E}[t]\}$, where $\mathcal{V}=\{1,...,n\}$ is the \emph{node set} and $\mathcal{E}[t] \subset \mathcal{V} \times \mathcal{V}$ is the \emph{directed edge set} at time $t$. The nodes are assumed to have unique identifiers that form a totally ordered set $\mathcal{I}$. Without loss of generality\footnote{There exists a bijection from $\mathcal{I}$ to $\mathcal{V}=\mathcal{N}\cup\mathcal{A}$.}, the node set is partitioned into a set of $N$ {\it normal nodes} $\mathcal{N}=\{1,2,\dotsc,N\}$ and a set of $A$ {\it adversary nodes} $\mathcal{A}=\{N+1,N+2,\dotsc,n\}$, with $A=n-N$. Let $\Gamma_n$ denote the set of all digraphs on $n$ nodes, which is of course a finite set. Note that $\mathcal{D}[t]\in\Gamma_n$ for all $t$, where $t\in\mathbb{R}_{\geq 0}$ for continuous time and $t\in\mathbb{Z}_{\geq 0}$ for discrete time. When we wish to refer to both discrete and continuous time, we generically say {\it at time $t$}.

The time-varying topology of the network is governed by a piecewise constant switching signal $\sigma(\cdot)$, which is defined on $\mathbb{Z}_{\geq0}$ for discrete time and $\mathbb{R}_{\geq0}$ for continuous time, and takes values in $\Gamma_n$. In order to emphasize the role of the switching signal, we denote $\mathcal{D}_{\sigma(t)}=\mathcal{D}[t]$. Let $\{\tau_{k}\}$, $k\in\mathbb{Z}_{\geq 0}$ denote the set of switching instances. For continuous time, we assume that there exists some constant $\tau \in \mathbb{R}_{>0}$ such that $\tau_{k+1}-\tau_k \geq \tau$ for all $k\geq0$. In other words, $\sigma(\cdot)$ is subject to the {\it dwell time} $\tau$. 

Each directed edge $(j,i)\in \mathcal{E}[t]$ models \emph{information flow} and indicates that node $i$ can be influenced by (or receive information from) node $j$ at time $t$. The set of \emph{in-neighbors}, or just \emph{neighbors}, of node $i$ at time $t$ is defined as $\mathcal{V}_i[t]=\{j\in\mathcal{V}\colon (j,i)\in \mathcal{E}[t]\}$ and the (in-)degree of $i$ is denoted $d_i[t] =\rvert \mathcal{V}_i[t] \rvert$. Likewise, the set of \emph{out-neighbors} of node $i$ at time $t$ is defined as $\mathcal{V}^{\text{out}}_i[t]=\{j\in\mathcal{V}\colon (i,j)\in \mathcal{E}[t]\}$. Because each node has access to its own state at time $t$, we also consider the \emph{inclusive neighbors} of node $i$, denoted $\mathcal{J}_i[t] = \mathcal{V}_i[t] \cup \{i\}$. Note that time-invariant networks are represented simply by dropping the dependence on time $t$.

\subsection{Update Model}
\label{S:UpdateModel}
Suppose that each node $i\in\mathcal{V}$ begins with some private value $x_i[0]\in \mathbb{R}$ (representing a measurement, opinion, vote, etc.), which evolves over time. Let $x_{\mathcal{N}}[t]=[x_1[t],x_2[t],\dotsc,x_N[t]]^\mathsf{T}$ and $x_{\mathcal{A}}[t]=[x_{N+1}[t],x_{N+2}[t],\dotsc,x_n[t]]^\mathsf{T}$ denote collectively the value (or state\footnote{Throughout this paper we refer to a node's value and state interchangeably.}) trajectories of the normal and adversary nodes, respectively, and let $x[t]=[x^\mathsf{T}_{\mathcal{N}}[t],x^\mathsf{T}_{\mathcal{A}}[t]]^{\mathsf{T}}$. The nodes interact synchronously by conveying their value to (out-)neighbors in the network. Each normal node updates its value over time according to a prescribed rule, which is modeled as 
\begin{equation*}
D \left[ x_i[t] \right] = f_{i,\sigma(t)}(t,x_{\mathcal{N}},x_{\mathcal{A}}), \enspace i \in \mathcal{N}, \mathcal{D}_{\sigma(t)}\in\Gamma_n,
\end{equation*}
where $D \left[x_i[t] \right]=\dot{x}_i[t]$ is the \emph{derivative operator} for continuous time and $D \left[x_i[t] \right]=x_i[t+1]-x_i[t]$ is the \emph{forward difference operator} for discrete time. Collectively, we define the system of normal nodes by
\begin{equation}
\label{E:CoopAgents}
D \left[ x_{\mathcal{N}}[t] \right] = f_{\sigma(t)}(t,x_{\mathcal{N}},x_{\mathcal{A}}), \enspace x_{\mathcal{N}}[0]\in\mathbb{R}^N, \mathcal{D}_{\sigma(t)}\in\Gamma_n,
\end{equation}
where $f_{\sigma(t)}(\cdot)=[f_{1,\sigma(t)}(\cdot),\dotsc,f_{N,\sigma(t)}(\cdot)]^{\mathsf{T}}$. Each of the functions $f_{i,\sigma(t)}(\cdot)$ can be arbitrary,\footnote{In continuous time, $f_{\sigma(t)}(\cdot)$ must satisfy appropriate assumptions to ensure existence of solutions.} and may be different for each node, depending on its role in the network.  These functions are designed {\it a priori} so that the normal nodes reach consensus. However, some of the nodes may not follow the prescribed strategy if they are compromised by an adversary. Such misbehaving nodes threaten the group objective, and it is important to design the $f_{i,\sigma(t)}(\cdot)$'s in such a way that the influence of such nodes can be eliminated or reduced without prior knowledge about their identities.

\subsection{Threat Model}
\begin{definition}
A node $k\in\mathcal{A}$ is said to be \textbf{malicious} if
\begin{itemize}
	\item it is not normal (i.e., it does not follow the prescribed update model either for at least one time-step in discrete time, or for some time interval of nonzero Lebesgue measure in continuous time);
	\item it conveys the same value, $x_k[t]$, to each out-neighbor;
	\item (for continuous-time systems) its value trajectory, $x_k[t]$ $\forall t$, is a uniformly continuous function of time on $[0,\infty)$.
\end{itemize}
\end{definition}

A few remarks are in order concerning malicious nodes. First, each malicious node is allowed to be omniscient (i.e., it knows all other values and the full network topology; it is aware of the update rules $f_{i,\sigma(t)}(\cdot)$, $\forall i\in\mathcal{N}$; it knows which other nodes are adversaries; and it knows the plans of the other adversaries).  The statement in the definition that the malicious nodes are not normal is intended to capture the idea that they do not apply the prescribed update rule for all time. The second assumption is intended as an assertion on the network realization. That is, if the network is realized through sensing or broadcast communication, it is assumed that the out-neighbors receive the same information. The third point is a technical assumption that applies only to malicious nodes modeled in continuous time. 
Limited only by these assumptions, the malicious nodes are otherwise allowed to operate in an arbitrary (potentially worst case) manner.

\subsection{Scope of Threats}
While there are various stochastic models that could be used to formalize the threat assumptions, here we use a deterministic approach and consider upper bounds on the number of compromised nodes either in the network ($F$-total) or in each node's neighborhood ($F$-local).

\begin{definition}[$F$-total set]
A set $\mathcal{S} \subset \mathcal{V}$ is \textbf{$F$-total} if it contains at most $F$ nodes in the network, i.e., $\rvert \mathcal{S}\rvert \le F$, $F\in\mathbb{Z}_{\geq0}$.
\label{def:f_local}
\end{definition}

\begin{definition}[$F$-local set]
A set $\mathcal{S} \subset \mathcal{V}$ is \textbf{$F$-local} if it contains at most $F$ nodes in the neighborhood of the other nodes for all $t$, i.e., $\rvert \mathcal{V}_i[t]\bigcap \mathcal{S}\rvert \le F$, $\forall i\in \mathcal{V}\setminus\mathcal{S}$, $F\in\mathbb{Z}_{\geq0}$.
\end{definition}

It should be noted that because the network topology may be time-varying, the local properties defining an $F$-local set must hold at all time instances. These definitions facilitate the definitions of the scope of threat models.

\begin{definition}
A set of adversary nodes is \textbf{$F$-totally bounded} or \textbf{$F$-locally bounded} if it is an $F$-total set or $F$-local set, respectively. We refer to these threat scopes as the \textbf{$F$-total} and \textbf{$F$-local} models, respectively.
\end{definition}

Note that whenever the set of $A$ adversary nodes $\mathcal{A}$ is $F$-totally bounded, we know $A\leq F$. On the other hand if $\mathcal{A}$ is $F$-locally bounded, it is possible that $A>F$. Indeed, there is no upper bound for $F$-locally bounded $\mathcal{A}$ since it is feasible that many adversaries may not be neighbors with any of the normal nodes over time. As a matter of terminology, we will refer to the threat model consisting of $F$-totally (or $F$-locally) bounded malicious nodes as the $F$-total malicious model (or $F$-local malicious model). The $F$-total fault model has been studied in distributed computing~\cite{Lamport82_ByzGeneralProb, lynch96:dist_algs, Vaidya2012_ItAppByzConsInArbDigraphs} and mobile robotics~\cite{Agmon06_FaultTolGathAlgs4AutMobRobots, Defago06_FTSSMobileRobGather, Bouzid2010_OptByzResConvInUniDimRobNets} for both stopping (or crash) failures and Byzantine failures. The $F$-local fault model has been studied in the context of Byzantine fault-tolerant broadcasting \cite{Pelc05_BroadWithLocalBoundByzanFaults, Ichimura10_ANewParaForBroad}.

\subsection{Resilient Asymptotic Consensus}
Given the threat model and scope of threats, we formally define resilient asymptotic consensus. 
Let $M[t]$ and $m[t]$ be the {\it maximum} and {\it minimum} values of the normal nodes at time $t$, respectively.
\begin{definition}[Resilient Asymptotic Consensus] 
\hspace{-0.15cm}The normal nodes are said to achieve \textbf{resilient asymptotic consensus} in the presence of $(a)$ $F$-totally bounded, or $(b)$ $F$-locally bounded 
misbehaving nodes if 
\begin{itemize}
\item $\exists L\in\mathbb{R}$ such that $\lim_{t\rightarrow\infty}x_i[t]=L$ for all $i\in\mathcal{N}$, and
\item $[m[0], M[0]]$ is an invariant set (i.e., the normal values remain in the interval for all $t$),
\end{itemize}
for any choice of initial values. Whenever the scope of threat is understood, we simply say that the normal nodes reach \textbf{asymptotic consensus}.
\end{definition}

The resilient asymptotic consensus problem has three important conditions. First, the normal nodes must reach asymptotic consensus in the presence of misbehaving nodes given a particular threat model (e.g., malicious) and scope of threat (e.g., $F$-total). This is a condition on {\it agreement}. Additionally, it is required that the interval containing the initial values of the normal nodes is an invariant set for the normal nodes; 
this is a {\it safety} condition. This safety condition is important when the current estimate of the consensus value is used in a safety critical process and the interval $[m[0], M[0]]$ is known to be safe. The agreement and safety conditions, when combined, imply a third condition on {\it validity}: the consensus quantity that the values of the normal nodes converge to must lie within the range of initial values of the normal nodes.

The validity condition is reasonable in applications where any value in the range of initial values of normal nodes is acceptable to select as the consensus value.  
For instance, consider a large sensor network where every sensor takes a measurement of its environment, captured as a real number.  Suppose that at the time of measurement, all values taken by correct sensors fall within a range $[a,b]$, and that all sensors are required to come to an agreement on a common measurement value. If the range of measurements taken by the normal sensors is relatively small, it will likely be the case that reaching agreement on a value within that range will form a reasonable estimate of the measurements taken by all sensors. However, if a set of malicious nodes is capable of biasing the consensus value outside of this range, the error in the measurements could be arbitrarily large.

More generally, suppose the nodes are trying to distributively minimize $\sum h_i(\theta)$, where each of the $h_i$'s is a local convex function and $\theta$ is the optimization variable. If the initial value of each node $i$ represents the value of $\theta$ that minimizes $h_i$, a convex combination of these initial values will represent an estimate of the optimal $\theta$, within some bounded error. On the other hand, if an adversary is capable of biasing the consensus value arbitrarily, the resulting value of the objective function will also be arbitrarily far away from its minimum value. One can formulate similar motivating examples for the validity condition in other applications as well; for instance, a swarm of robots that are trying to flock should not be pulled in arbitrary directions by a malicious agent in the network.  

\section{\uppercase{Consensus Algorithm}}
\label{S:Algorithm}
Linear consensus algorithms have attracted significant interest in recent years~\cite{OlfatiSaber07_ConsensusAndCoopInNetMultiagentSys, Ren07_InfoConsInMultiVehCoopCtrl}, due to their applicability in a variety of contexts. In such strategies, at time $t$, each node senses or receives information from its neighbors, and changes its value according to
\begin{equation}
\label{E:LCP}
D[x_i[t]]=\sum_{j\in \mathcal{J}_i[t]}w_{ij}[t]x_j[t],
\end{equation}
where $w_{ij}[t]$ is the weight assigned to node $j$'s value by node $i$ at time $t$.

Different conditions have been reported in the literature to ensure asymptotic consensus is reached~\cite{Xiao04fastlinear,Ren05_ConsSeekInMultiAgentSysUnderDynChangIntTop,Moreau05_StabOfMultiAgSysWithTimeDepCommLinks,Jadbabaie03_CoordOfGroupsOfMobileAutAgentsUsingNearNeighRule,Tsitsiklis1984_ProbsInDecentDecMakComp}.  In discrete time, it is common to assume that there exists a constant $\alpha \in \mathbb{R}$, $0 < \alpha < 1$ such that all of the following conditions hold:\footnote{The conditions on the weights are modified from what is reported in the literature to account for the forward difference operator. Accounting for this, the updated value of each node is formed as a convex combination of the neighboring values and its own value.}
\begin{itemize}
\item $w_{ij}[t]=0$ whenever $j\not\in \mathcal{J}_i[t], i\in\mathcal{N}$, $t\in\mathbb{Z}_{\geq0}$;
\item $w_{ij}[t] \ge\alpha$,  $\forall j\in \mathcal{V}_i[t], i\in\mathcal{N}$, $t\in\mathbb{Z}_{\geq0}$;
\item $w_{ii}[t] \ge\alpha-1$, $\forall i\in\mathcal{N}$, $t\in\mathbb{Z}_{\geq0}$; 
\item $\sum_{j=1}^{n}w_{ij}[t]=0$,  $\forall i\in\mathcal{N}$, $t\in\mathbb{Z}_{\geq0}$. 
\end{itemize}

In continuous time there are similar conditions, except in this case the self-weights are given by
\begin{equation*}
w_{ii}[t]=-\sum_{j \in \mathcal{V}_i[t]} w_{ij}[t], \quad \forall i\in\mathcal{N}, \forall t\in\mathbb{R}_{\geq0}.
\end{equation*}
In this case, the weights must be piecewise continuous and uniformly bounded. That is, there exists $\beta \in \mathbb{R}_{>0}$, $\beta\geq\alpha$, such that $w_{ij}[t]\leq \beta$, for all $i,j\in\mathcal{N}$ and $t\in\mathbb{R}_{\geq0}$. Similar to the discrete time case, the weights $w_{ij}[t]$ are zero precisely whenever $j\not\in \mathcal{J}_i[t]$, and bounded below by $\alpha$ otherwise.
Together, these conditions imply the analogue of the fourth condition above.

Given these conditions, a necessary and sufficient condition for reaching asymptotic consensus in time-invariant networks is that the digraph has a \emph{rooted out-branching}, also called a \emph{rooted directed spanning tree}~\cite{Ren07_InfoConsInMultiVehCoopCtrl}. The case of dynamic networks is not quite as straightforward. In this case, under the conditions stated above, a sufficient condition for reaching asymptotic consensus is that there exists a uniformly bounded sequence of contiguous time intervals such that the union of digraphs across each interval has a rooted out-branching~\cite{Ren05_ConsSeekInMultiAgentSysUnderDynChangIntTop}. Recently, a more general condition referred to as the {\it infinite flow property} has been shown to be both necessary and sufficient for asymptotic consensus for a class of discrete-time stochastic models~\cite{Touri2011_OnErgodicityInfFlowCons}. 
Finally, the lower bound on the weights is needed because there are examples of asymptotically vanishing weights in which consensus is not reached \cite{Lorenz2010_OnCond4Conv2Cons}.  

In general, the problem of selecting the best weights in the linear update rule (\ref{E:LCP}) is nontrivial, and the choice affects the rate of consensus. The problem of selecting the optimal weights (with respect to the speed of the consensus process) in time-invariant, discrete-time, bidirectional networks is addressed in \cite{Xiao04fastlinear} by formulating a semidefinite program (SDP). However, this SDP is solved at design time with global knowledge of the network topology. A simple choice of weights for discrete-time systems that requires only local information is to let $w_{ij}[t]=1/(1+d_i[t])$ for $j\in\mathcal{V}_i[t]$ and $w_{ii}[t]=-d_i[t]/(1+d_i[t])$. In continuous time, a simple choice is to let $w_{ij}\equiv 1$ for $j\in\mathcal{V}_i[t]$ and $w_{ii}[t]=-d_i[t]$.

One problem with the linear update given in (\ref{E:LCP}) is that it is not resilient to misbehaving nodes. In fact, it was shown in \cite{Jadbabaie03_CoordOfGroupsOfMobileAutAgentsUsingNearNeighRule, Gupta06_OnRobustnessOfDistAlgs} that a single `leader' node can cause all agents to reach consensus on an arbitrary value of its choosing (potentially resulting in a dangerous situation in physical systems).

The Adversarial Robust Consensus Protocol (ARC-P) addresses this vulnerability of the linear update of (\ref{E:LCP}) by a simple modification. Instead of trusting every neighbor by using every value in the update, the normal node first removes the extreme values from consideration in the update by effectively setting their weights (temporarily) to zero. It is be shown in subsequent sections that this simple strategy provides resilience against malicious nodes in robust networks. 

\subsection{Description of ARC-P}
At time $t$, each normal node $i$ obtains the values of other nodes in its neighborhood. At most $F$ of node $i$'s neighbors may be malicious; however, node $i$ is unsure of which neighbors may be compromised. To ensure that node $i$ updates its value in a safe manner, it removes the extreme values with respect to its own value according to the following protocol. 

\begin{enumerate}
\item At time $t$, each normal node $i$ obtains the values of its neighbors, and forms a sorted list. 
\item If there are less than $F$ values strictly larger than its own value, $x_i[t]$, then normal node $i$ removes all values that are strictly larger than its own. Otherwise, it removes precisely the largest $F$ values in the sorted list (breaking ties in a deterministic manner; e.g., by keeping the values of the nodes with the smaller unique identifiers in $\mathcal{I}$).  Likewise, if there are less than $F$ values strictly smaller than its own value, then node $i$ removes all values that are strictly smaller than its own. Otherwise, it removes precisely the smallest $F$ values. 
\item Let $\mathcal{R}_i[t]$ denote the set of nodes whose values were removed by normal node $i$ in step 2 at time $t$. Each normal node $i$ applies the update
\begin{equation}
D[x_i[t]]=\sum_{j\in \mathcal{J}_i[t]\setminus\mathcal{R}_i[t]}w_{ij}[t]x_j[t],
\label{eqn:ft_update}
\end{equation}
where the weights $w_{ij}[t]$ satisfy the conditions stated above, but with $\mathcal{J}_i[t]$ replaced by $\mathcal{J}_i[t]\setminus\mathcal{R}_i[t]$.\footnote{In this case, a simple choice for the weights in discrete time is to let $w_{ij}[t]=1/(1+d_i[t]-|\mathcal{R}_i[t]|)$ for $j\in\mathcal{V}_i[t]$ and $w_{ii}[t]=(|\mathcal{R}_i[t]|-d_i[t])/(1+d_i[t]-|\mathcal{R}_i[t]|)$. In continuous time, let $w_{ij}\equiv 1$ for $j\in\mathcal{V}_i[t]$ and $w_{ii}[t]=|\mathcal{R}_i[t]|-d_i[t]$.} Note that if all neighboring values are removed, then $D[x_i[t]]=0$. 
\end{enumerate}

As a matter of terminology, we refer to the bound on the number of larger or smaller values that could be thrown away as the {\it parameter} of the algorithm. 
Above, the parameter of ARC-P under the $F$-local and $F$-total models is $F$.

Observe that the set of nodes removed by normal node $i$, $\mathcal{R}_i[t]$, is possibly time-varying. Hence, even though the underlying network topology may be fixed, ARC-P effectively induces switching behavior, and can be viewed as the linear update of (\ref{E:LCP}) with a specific rule for state-dependent switching (the rule given in step 2).

\subsection{ARC-P in Continuous Time}
The previous section outlined the steps taken in ARC-P to remove the influence of nodes with extreme values. In order to analyze (\ref{E:CoopAgents}) for existence and uniqueness of solutions in continuous time, it is useful to express ARC-P as a composition of functions. For this, we require the following definitions.

\begin{definition}
\label{D:Functions}
Let $k \in \mathbb{N}$ and $F \in \mathbb{Z}_{\geq0}$. Denote the elements of vectors $\xi, w, z \in \mathbb{R}^{k}$ by $\xi_l$, $w_l$, and $z_l$, respectively, for $l=1, 2, \dotsc, k$. Then: 
\begin{enumerate}[(i)]
\item The (ascending) \textbf{sorting function} on $k$ elements, $\rho_{k} \colon \mathbb{R}^{k} \rightarrow \mathbb{R}^{k}$, is defined by $\xi = \rho_{k}(z)$ such that $\xi$ is a permutation of $z$ which satisfies
\begin{equation}
\label{E:SortAscending}    
\xi_{1} \leq \xi_{2} \leq \dots \leq \xi_{k};
\end{equation}
\item The \textbf{weighted zero-selective reduce function} with respect to $F$ and $k$, $r_{0,F}^k \colon \mathbb{R}^k \times \mathbb{R}^{k} \rightarrow \mathbb{R}$, is defined by (\ref{E:WeightedZeroReduce}), where $1_{\geq 0}(\alpha)$ and $1_{\leq 0}(\alpha)$ are indicator functions, and the weights are uniformly bounded by $0<\alpha\leq w_l \leq\beta$, $\forall l$.
\begin{figure*}[!t]
\normalsize
\begin{equation}
\label{E:WeightedZeroReduce}
r_{0,F}^k (z,w) = \begin{cases}
\sum_{l=1}^{F} w_l 1_{\geq 0}(z_l)z_l+ \sum_{l=F+1}^{k-F} w_l z_l + \sum_{l=k-F+1}^{k} w_l 1_{\leq 0}(z_l)z_l & \text{$k>2F$;} \\
\sum_{l=1}^{k-F} w_l 1_{\geq 0}(z_l)z_l + \sum_{l=F+1}^{k} w_l 1_{\leq 0}(z_l)z_l  & \text{$F<k\leq 2F$;} \\
0 & \text{$k\leq F$;}
\end{cases}
\end{equation}
\hrulefill
\vspace*{4pt}
\end{figure*}
\item The composition of the sorting and weighted zero-selective reduce functions with respect to $F$ and $k$ is defined by $\phi_F^k \colon \mathbb{R}^{k} \times \mathbb{R}^{k} \rightarrow \mathbb{R}$, which is defined for all $z \in \mathbb{R}^{k}$ and $w \in \mathbb{R}^k$ such that $0<\alpha\leq w_l \leq\beta$ by
\begin{equation*}
\label{E:Compensator}
\phi_F^k(z,w)= r_{0,F}^k (\rho_{k}(z),w). 
\end{equation*} 
\end{enumerate}
\end{definition}

Then, the update rule of ARC-P for each normal node $i \in \mathcal{N}$ for $t\in\mathbb{R}_{\geq0}$ is given by
\begin{equation}
\label{E:RobustConsProtocol}
f_{i,\sigma(t)}(t,x_{\mathcal{N}},x_{\mathcal{A}})=\phi_{F}^{d_i[t]}\left(J_i[t](x[t]-x_i[t] 1_n), w_i[t] \right), 
\end{equation}
in which $x[t]=[x_{\mathcal{N}}^{\mathsf{T}}[t], x_{\mathcal{A}}^{\mathsf{T}}[t]]^{\mathsf{T}} \in \mathbb{R}^n$ and $1_n \in \mathbb{R}^n$ is the vector of ones. The time-varying weight vector 
$$
w_i[t]=[w_{ii_1[t]}[t],w_{ii_2[t]}[t],\dotsc,w_{ii_{d_i[t]}[t]}[t]]^{\mathsf{T}}, 
$$
satisfies the bound $0<\alpha\leq w_{ii_j[t]} \leq\beta$ for all $j=1,2,\dotsc,d_i[t]$, where $i_1[t], i_2[t], \dotsc, i_{d_i[t]}[t]$ are the node indices of the neighbors of node $i$ in the order determined by the sorting function at time $t$ (i.e., according to (\ref{E:SortAscending}) such that the weights match the corresponding neighbor). Finally, $J_i[t] \in \mathbb{R}^{(d_i[t]) \times n}$ is a sparse matrix with each row corresponding to a distinct $j\in \mathcal{V}_i[t]$ such that each row has a single $1$ in the $j$-th column. Thus, there is a one-to-one correspondence between $j\in \mathcal{V}_i[t]$ and rows in $J_i[t]$. These terms are defined so that (\ref{E:RobustConsProtocol}) is equivalent to (\ref{eqn:ft_update}) for all $t\in\mathbb{R}_{\geq0}$.

\subsubsection{Existence and Uniqueness of Solutions}
As a first step toward showing existence and uniqueness of solutions, we show that (\ref{E:RobustConsProtocol}) satisfies a Lipschitz condition for all $i\in\mathcal{N}$.
\begin{definition}
\label{D:LipschitzCty}
Let $||\cdot ||$ denote any norm defined on a Euclidean space, and let $g(t,x,u)$, $g \colon \mathbb{R} \times \mathbb{R}^n \times \mathbb{R}^p \rightarrow \mathbb{R}^q$, be a piecewise continuous function in $t$ and $u$. Then $g$ satisfies a {\it global Lipschitz condition} with \textit{Lipschitz constant} $L$ if the following condition holds for all $z,y \in \mathbb{R}^n$, $t\in\mathbb{R}$:
\begin{equation*}
\label{E:LipschitzCty}
||g(t,z,u)-g(t,y,u)|| \leq L ||z-y||.
\end{equation*}
\end{definition}

\begin{theorem}
\label{T:MainCtyResult}
The function $f_{\sigma(t)}(t,x_{\mathcal{N}},x_{\mathcal{A}})=f_{\sigma(t)}(t,x)$ that defines the dynamics of the normal nodes, with $f_{i,\sigma(t)}(\cdot)$ defined in (\ref{E:RobustConsProtocol}), satisfies a global Lipschitz condition in $x_\mathcal{N}$ and $x$.
\end{theorem}
\begin{IEEEproof} Because the weights are piecewise continuous and the switching signal is piecewise constant, it follows that $f_{\sigma(t)}(t,x)$ is piecewise continuous in $t$.  We first show that $f_{\sigma(t)}(t,x)$ satisfies a Lipschitz condition in $x$ by showing that the component functions $f_{i,\sigma(t)}(t,x)$ do. For this, fix $t\in\mathbb{R}_{\geq0}$, $F\in\mathbb{Z}_{\geq0}$, $d_i[t]=k$, and $w_i[t]=w$. The argument to $\phi_F^k(\cdot,w)$ is linear and the sorting function is Lipschitz, as shown in \cite{LeBlanc11_ConsNetMASWithAdv}. Hence, all there is to show is that the weighted zero-selective reduce function with respect to $F$ and $k$ is Lipschitz. Fix $z,y \in \mathbb{R}^k$. The key observation is that 
\begin{equation*}
1_{\geq 0}(z_l)z_l - 1_{\geq 0}(y_l)y_l \leq |z_l-y_l|,
\end{equation*}
for each $l=1,2,\dotsc,k$, which is trivial to show by checking the four cases depending on the signs of $z_l$ and $y_l$. Since $0<\alpha\leq w_l \leq\beta$, it follows that
\begin{equation*}
w_l 1_{\geq 0}(z_l)z_l - w_l 1_{\geq 0}(y_l)y_l \leq \beta|z_l-y_l|,
\end{equation*}
Likewise, the inequality holds when the indicator function is $1_{\leq 0}(\cdot)$ instead of $1_{\geq 0}(\cdot)$. Combining this with the triangle inequality, it is straightforward to show using the Manhattan norm that $r_{0,F}^k$ is Lipschitz with Lipschitz constant $\beta$. Finally, we show $f_{\sigma(t)}(t,x_{\mathcal{N}},x_{\mathcal{A}})$ satisfies a Lipschitz condition in $x_{\mathcal{N}}$. Fix $y,z \in \mathbb{R}^N$ and note that the malicious nodes' trajectories are uniformly continuous in time by assumption (and therefore $f_{\sigma(t)}(t,x_{\mathcal{N}},x_{\mathcal{A}})$ is piecewise continuous in time). Since, there exists a global Lipschitz constant for $x$, denoted $L$, we know
\begin{align*}
||f_{\sigma(t)}(t,y,x_{\mathcal{A}})&-f_{\sigma(t)}(t,z,x_{\mathcal{A}})||\\
&\leq L\left|\left|\left[ \begin{array}{c}y\\x_{\mathcal{A}} \end{array} \right]-\left[ \begin{array}{c}z\\x_{\mathcal{A}} \end{array} \right]\right|\right|=L||y-z||.
\end{align*}
\end{IEEEproof}

Since we assume that $\sigma(t)$ is piecewise constant, $x_{\mathcal{A}}$ is piecewise continuous (in fact we assume it is uniformly continuous on $[0,\infty)$), and the weights are piecewise continuous, it follows that $f_{\sigma(t)}(t,x_{\mathcal{N}},x_{\mathcal{A}})$ defined by (\ref{E:CoopAgents}) with component functions given in (\ref{E:RobustConsProtocol}) is piecewise continuous in $t$. Theorem \ref{T:MainCtyResult} shows that $f_{\sigma(t)}(\cdot)$ is Lipschitz in $x_{\mathcal{N}}$. We show next in Lemma \ref{L:BoundAveragePhi} that $f_{\sigma(t)}(\cdot)$ is bounded by the current normal values $x_{\mathcal{N}}[t]$ for $t\in\mathbb{R}_{\geq0}$. From these facts, we conclude the local existence and uniqueness of solutions of (\ref{E:RobustConsProtocol}) for all $i\in\mathcal{N}$. Then, we show in Lemma \ref{L:HypercubeInv} that any solution is confined to a compact set, from which we conclude global existence and uniqueness of solutions of (\ref{E:RobustConsProtocol}) for all $i\in\mathcal{N}$.

\begin{lemma}
\label{L:BoundAveragePhi}
Consider the normal node $i \in \mathcal{N}$ with continuous dynamics executing ARC-P with parameter $F\in\mathbb{Z}_{\geq0}$ and assume there are at most $F$ adversary nodes in its neighborhood at time $t$. Then, for each $t \in \mathbb{R}_{\geq0}$ 
\begin{equation*}
B(m[t]-x_i[t]) \leq f_{i,\sigma(t)}(x_{\mathcal{N}}, x_{\mathcal{A}}) \leq B(M[t]-x_i[t]),
\end{equation*}
where $B=\beta(n-F-1)$, $m[t]=\min_{j\in\mathcal{N}}\{x_j[t]\}$, and $M[t]=\max_{k\in\mathcal{N}}\{x_k[t]\}$.
\end{lemma}
\begin{IEEEproof}
If $d_i[t]\leq F$, or if $F<d_i[t]\leq2F$ and there are at most $F$ neighbors with larger and smaller values than $x_i[t]$, then $f_{i,\sigma(t)}(t,x_{\mathcal{N}}, x_{\mathcal{A}})=0$, and the result follows. Therefore, assume $d_i[t]>F$ and at least one value not equal to $x_i[t]$ is used in the update at time $t$, say $x_j[t]$. Suppose $x_j[t]>M[t]$. Then, by definition $j$ must be an adversary and $x_j[t]>x_i[t]$. Since $i$ uses $x_j[t]$ at time $t$, there must be at least $F$ more nodes in the neighborhood of $i$ with values at least as large as $x_j[t]$. Hence, these nodes must also be adversaries, which contradicts the assumption of at most $F$ adversary nodes in the neighborhood of $i$ at time $t$. Thus, $x_j[t]\leq M[t]$. Similarly, we can show that $x_j[t]\geq m[t]$. By combining the fact that there are at most $n-1$ neighbors of $i$, at least $F$ values will be removed (since $d_i[t]>F$), and $w_{ij}[t]\leq\beta$ for all $j\in\mathcal{V}_i[t]$, it follows that
$$
B(m[t]-x_i[t])\leq \hspace{-0.3cm}\sum_{j\in\mathcal{V}_i[t]\setminus\mathcal{R}_i[t]}\hspace{-0.3cm}w_{ij}[t](x_j[t]-x_i[t])\leq B(M[t]-x_i[t]).
$$ 
\end{IEEEproof}

Observe that Lemma \ref{L:BoundAveragePhi} holds under both the $F$-total and $F$-local models, and bounds $f_{\sigma(t)}(\cdot)$ as a function of the total number of nodes $n$, the upper bound on the number of adversaries in the neighborhood of any normal node $F$, and the current state of the normal node values $x_{\mathcal{N}}[t]$. The next result shows that for any solution of (\ref{E:CoopAgents}), the hypercube $\mathcal{H}_0$, which is given by $[m[0],M[0]]^N$, is a \emph{robustly positively invariant set} (defined as follows). 
\begin{definition}
\label{D:RobustPosInvSet}
The set $\mathcal{S} \subset \mathbb{R}^N$ is \textbf{robustly positively invariant} for the system given by (\ref{E:CoopAgents})
if for all $x_{\mathcal{N}}[0] \in \mathcal{S}$, $x_{\mathcal{A}}[t] \in \mathbb{R}^A$, any solution satisfies $x_{\mathcal{N}}[t] \in \mathcal{S}$ for all $t\geq0$.
\end{definition}

\begin{lemma}
\label{L:HypercubeInv}
Suppose the normal nodes in $\mathcal{N}$ have continuous dynamics and use ARC-P with parameter $F\in\mathbb{Z}_{\geq0}$ under the $F$-local or $F$-total malicious model. Then, the hypercube $\mathcal{H}_0 = [m[0],M[0]]^N$ defined by
\begin{equation*}
\label{E:Hypercube}
\mathcal{H}_0=\{ y \in \mathbb{R}^N \colon m[0] \leq y_i \leq M[0], \ i=1,2,\dotsc,N \},
\end{equation*}
is robustly positively invariant for the system of normal nodes.
\end{lemma}
\begin{IEEEproof}
Since $\mathcal{H}_0$ is compact and any solution of (\ref{E:CoopAgents}) using (\ref{E:RobustConsProtocol}) is continuous with $x_{\mathcal{N}}[0] \in \mathcal{H}_0$, we must show that $f_{\sigma(t)}(\cdot)$ is not directed outside of $\mathcal{H}_0$, whenever $x_{\mathcal{N}}[t] \in \partial \mathcal{H}_0$, for all $\mathcal{D}_{\sigma(t)}\in\Gamma_n$ and all allowable trajectories of $x_{\mathcal{A}}$. 
The boundary of $\mathcal{H}_0$ is given by
\begin{equation*}
\partial \mathcal{H}_0=\{ y \in \mathcal{H}_0 \colon \exists i \in \{1,2,\dotsc , N \} \text{ s.t. } y_i \in \{m[0], M[0] \} \}.
\end{equation*}

Now, fix $x_{\mathcal{N}} \in \partial \mathcal{H}_0$ for some $t\in\mathbb{R}_{\geq0}$. Let $e_j$ denote the $j$-th canonical basis vector and denote $\mathcal{I}_{\mathcal{N},\text{min}}, \mathcal{I}_{\mathcal{N},\text{max}} \subseteq \{1,2,\dotsc, N \}$ as the sets defined by
\begin{equation*}
j \in \mathcal{I}_{\mathcal{N},\text{min}} \Leftrightarrow x_j = m[0] \text{ and } k \in \mathcal{I}_{\mathcal{N},\text{max}} \Leftrightarrow x_k = M[0].
\end{equation*}
Then, from the geometry of the hypercube, we require
\begin{align*}
e_j^\mathsf{T} f_{\sigma(t)}(t, x_{\mathcal{N}},x_{\mathcal{A}}) &\geq 0 \quad \forall j \in \mathcal{I}_{\mathcal{N},\text{min}}, \\
e_k^\mathsf{T} f_{\sigma(t)}(t, x_{\mathcal{N}},x_{\mathcal{A}}) &\leq 0 \quad \forall k \in \mathcal{I}_{\mathcal{N},\text{max}}.
\end{align*}
These conditions are true for all $\mathcal{D}_{\sigma(t)}\in\Gamma_n$and $x_{\mathcal{A}}$ under the $F$-local or $F$-total models by Lemma~\ref{L:BoundAveragePhi}, in which the lower bound is used for $j \in \mathcal{I}_{\mathcal{N},\text{min}}$ (since $x_j = m[0]$), and the upper bound is used for $k \in \mathcal{I}_{\mathcal{N},\text{max}}$ (since $x_k = M[0]$). 
\end{IEEEproof}

The argument made in Lemma~\ref{L:BoundAveragePhi} implies that any time an adversary under the $F$-total or $F$-local model is outside of  $\mathcal{I}_t=[m[t],M[t]]$, its influence is guaranteed to be removed by its normal neighbors, and therefore has the same effect as if it were on the boundary of $\mathcal{I}_t$. Using Lemma~\ref{L:HypercubeInv} we conclude $\mathcal{I}_t\subseteq\mathcal{I}_0$, $\forall t\geq0$. Hence, each adversary is effectively restricted to the compact set $\mathcal{I}_0$, with respect to (\ref{E:CoopAgents}). This fact enables us to allow adversary states in $\mathbb{R}^A$ rather than explicitly restricting them to a compact set, while still ensuring existence and uniqueness of solutions.
\begin{corollary}
\label{C:Uniqueness}
Given the choice of bounded, piecewise continuous, time-varying weights, piecewise constant switching signal, and adversaries (i.e., adversary value trajectories) that satisfy the $F$-local or $F$-total malicious model, the system of normal nodes defined by (\ref{E:CoopAgents}) with component functions given in (\ref{E:RobustConsProtocol}) has a unique solution for all $t \geq 0$ and for any $x_{\mathcal{N}}[0] \in \mathbb{R}^N$.
\end{corollary}


\section{\uppercase{Robust Network Topologies}}
\label{S:Topologies}
\subsection{Network Robustness}
In this section, we introduce {\it robust network topologies} that satisfy certain graph theoretic properties, which we refer to generically as {\it network robustness}. Network robustness formalizes the notion of sufficient redundancy of information flow to subsets of a network in a single hop. Therefore, this property holds promise to be effective for the study of resilient distributed algorithms that use only local information. In contrast, network connectivity formalizes the notion of sufficient redundancy of information flow across the network through independent paths. Due to the fact that each independent path may include multiple intermediate nodes, network connectivity is well-suited for studying resilient distributed algorithms that assume such nonlocal information is available (for example, by explicitly relaying information across multiple hops in the network \cite{lynch96:dist_algs}, or by `inverting' the dynamics on the network to recover the needed information \cite{Sundaram2011_DistFunctCalcViaLinItStratInPresOfMalAgs,Pasqualetti2011_ConsCompInUnrelNets}). However, network connectivity is no longer an appropriate metric for an algorithm that uses purely local information, such as ARC-P. This is demonstrated by the following proposition~\cite{Zhang2012_RobInfoDiffAlgs2LocBdAdv}.

\begin{proposition}
There exists a graph with connectivity $\kappa = \lfloor \frac{n}{2}\rfloor +F -1 $ in which ARC-P does not ensure asymptotic consensus.
\label{prop:counter_example}
\end{proposition}

\begin{figure}
\centering
\includegraphics[width=7.5cm]{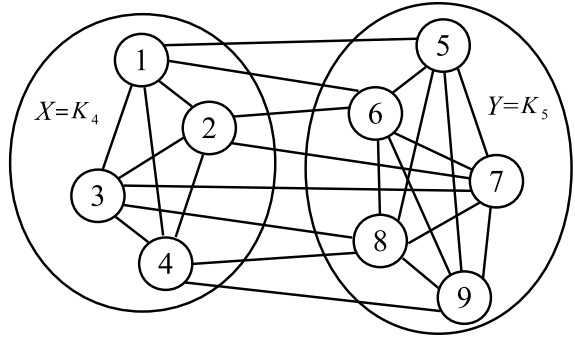}
\caption{Example of a $5$-connected graph satisfying Prop.~\ref{prop:counter_example} whenever $F=2$.}
\label{fig:Counterexample}
\end{figure}

Figure \ref{fig:Counterexample} illustrates an example of this kind of graph with $n=9$, $F=2$, and $\kappa=5$. In this graph, there are two cliques (complete subgraphs), $X=K_4$ and $Y=K_5$, where $K_n$ is the complete graph on $n$ nodes. Each node in $X$ has exactly $F=2$ neighbors in $Y$, and all but two nodes in $Y$ have $F=2$ neighbors in $X$ (nodes 5 and 9 have only one neighbor in $X$, because otherwise a node in $X$ would have more than $F=2$ neighbors in $Y$). One can see that if the initial values of nodes in $X$ and $Y$ are $a\in\mathbb{R}$ and $b\in\mathbb{R}$, respectively, with $a\neq b$, then asymptotic consensus is not achieved whenever ARC-P is used with parameter $F$ -- even in the absence of misbehaving nodes.  This is because each node views the values of its $F$ neighbors from the opposing set as extreme, and removes all of these values from its list.  The only remaining values for each node are from its own set, and thus no node ever changes its value.

The situation can be even worse in the more general case of digraphs. Examples of digraphs are illustrated in \cite{LeBlanc_LowCompResConsAdv_HSCC12} that are $(n-1)$-connected and have minimum out-degree $n-2$, yet ARC-P still cannot guarantee asymptotic consensus. Thus, even digraphs with a relatively large connectivity (or minimum out-degree) are not sufficient to guarantee consensus of the normal nodes, indicating the inadequacy of these traditional metrics to analyze the convergence  properties of ARC-P. Taking a closer look at the graph in Fig.~\ref{fig:Counterexample}, we see that the reason for the failure of consensus is that no node has enough neighbors in the opposite set; this causes every node to throw away all useful information from outside of its set, and prevents consensus.  Based on this intuition, the following properties, i.e., $r$-reachable sets and $r$-robustness, were introduced in \cite{Zhang2012_RobInfoDiffAlgs2LocBdAdv}. 

\begin{definition}[$r$-reachable set]
Given a digraph $\mathcal{D}$ and a nonempty subset $\mathcal{S}$ of nodes of $\mathcal{D}$, we say $\mathcal{S}$ is an \textbf{$r$-reachable set} if $\exists i\in \mathcal{S}$ such that $\rvert \mathcal{V}_i\setminus\mathcal{S}\rvert \ge r$, where $r\in\mathbb{Z}_{\ge 0}$.
\end{definition}

A set $\mathcal{S}$ is $r$-reachable if it contains a node that has at least $r$ neighbors outside of $\mathcal{S}$. The parameter $r$ quantifies the redundancy of information flow from nodes outside of $\mathcal{S}$ to {\it some} node inside $\mathcal{S}$. Intuitively, the $r$-reachability property captures the idea that some node inside the set is influenced by a sufficiently large number of nodes from outside the set. The above reachability property pertains to a given set $\mathcal{S}$; in order to generalize this notion of redundancy to the entire network, we introduce the following definition of $r$-robustness.

\begin{definition}[$r$-robustness]
A nonempty, nontrivial digraph $\mathcal{D}=\{\mathcal{V},\mathcal{E}\}$ on $n$ nodes ($n\geq2$) is \textbf{$r$-robust}, with $r\in\mathbb{Z}_{\geq0}$, if for every pair of nonempty, disjoint subsets of $\mathcal{V}$, at least one of the subsets is $r$-reachable. By convention, if $\mathcal{D}$ is empty or trivial ($n\leq1$), then $\mathcal{D}$ is 0-robust. The trivial graph is also 1-robust.
\label{def:r_robust}
\end{definition}

The reason that pairs of nonempty, disjoint subsets of nodes are considered in the definition of $r$-robustness can be seen in the example of Fig.~\ref{fig:Counterexample}. If either $X$ or $Y$ were $3$-reachable ($r=F+1=3$), then at least one node would be sufficiently influenced by a node outside of its set in order to drive it away from the values of its group, and thereby lead its group to the values of the other set. However, if there are misbehaving nodes in the network, then the situation becomes more complex.  For example, consider the $F$-total model of malicious nodes, and consider two sets $X$ and $Y$ in the graph.  Let $s$ be the total number of nodes in these two sets that each have at least $F+1$ neighbors outside their own set.  If $s \le F$, then simply by choosing these nodes to be malicious, the sets $X$ and $Y$ contain no normal nodes that bring in enough information from outside, and thus the system can be prevented from reaching consensus.  This reasoning suggests a need to specify a minimum number of nodes that are sufficiently influenced from outside of their set (in this example, at least $F+1$ nodes). This intuition leads to the following generalizations of $r$-reachability and $r$-robustness.

\begin{definition}[($r,s$)-reachable set]
\label{D:RSReachable}
Given a digraph $\mathcal{D}$ and a nonempty subset of nodes $\mathcal{S}$, we say that $\mathcal{S}$ is an \textbf{$(r,s)$-reachable set} if there are at least $s$ nodes in $\mathcal{S}$ with at least $r$ neighbors outside of $\mathcal{S}$, where $r,s\in \mathbb{Z}_{\geq0}$; i.e., given $\mathcal{X}_{\mathcal{S}}=\{i\in \mathcal{S} \colon |\mathcal{V}_i \setminus \mathcal{S}| \geq r \}$, then $|\mathcal{X}_{\mathcal{S}}|\geq s$.
\end{definition}

Observe that $r$-reachability is equivalent to $(r,1)$-reachability; hence, $(r,s)$-reachability is a strict generalization of $r$-reachability. If a set $\mathcal{S}$ is $(r,s)$-reachable, we know there are at least $s$ nodes in $\mathcal{S}$ with at least $r$ neighbors outside of $\mathcal{S}$. Thus, if $\mathcal{S}$ is $(r,s)$-reachable, then it is $(r,s')$-reachable, for $s'\leq s$. Also, it is clear that $s\leq |\mathcal{S}|$ and all subsets of nodes of any digraph are $(r,0)$-reachable. The additional specificity on the number of nodes with redundant information flow from outside of their set is useful for defining a more general notion of robustness.

\begin{definition}[($r,s$)-robustness]
\label{D:RTotalRobust}
A nonempty, nontrivial digraph $\mathcal{D}$ $=$ $\{\mathcal{V},\mathcal{E}\}$ on $n$ nodes ($n\geq2$) is \textbf{$(r,s)$-robust}, for nonnegative integers $r\in\mathbb{Z}_{\geq0}$, $1\leq s\leq n$, if for every pair of nonempty, disjoint subsets $\mathcal{S}_1$ and $\mathcal{S}_2$ of $\mathcal{V}$ such that $\mathcal{S}_1$ is $(r,s_{r,1})$-reachable and $\mathcal{S}_2$ is $(r,s_{r,2})$-reachable with $s_{r,1}$ and $s_{r,2}$ maximal (i.e., $s_{r,k}=|\mathcal{X}_{\mathcal{S}_k}|$ where $\mathcal{X}_{\mathcal{S}_k}=\{i\in \mathcal{S}_k \colon |\mathcal{V}_i \setminus \mathcal{S}_k| \geq r \}$ for $k\in\{1,2\}$), 
then at least one of the following hold:
\begin{enumerate}[(i)]
\item $s_{r,1}=|\mathcal{S}_1|$;
\item $s_{r,2}=|\mathcal{S}_2|$;
\item $s_{r,1}+s_{r,2}\geq s$.
\end{enumerate}
By convention, if $\mathcal{D}$ is empty or trivial ($n\leq1$), then $\mathcal{D}$ is (0,1)-robust. If $\mathcal{D}$ is trivial, $\mathcal{D}$ is also (1,1)-robust.
\end{definition}

A few remarks are in order with respect to this definition. The definition of ($r,s$)-robustness aims to capture the idea that enough nodes in every pair of nonempty, disjoint sets $\mathcal{S}_1,\mathcal{S}_2 \subset \mathcal{V}$ have at least $r$ neighbors outside of their respective sets. To quantify what is meant by ``enough'' nodes, it is necessary to take the maximal $s_{r,k}$ for which $\mathcal{S}_k$ is $(r,s_{r,k})$-reachable for $k\in \{1,2\}$ (since $\mathcal{S}_k$ is $(r,s_{r,k}')$-reachable for $s_{r,k}'\leq s_{r,k}$). Since $s_{r,k}=|\mathcal{X}_{\mathcal{S}_k}|$, condition $(i)$ or $(ii)$ means that {\it all} nodes in $\mathcal{S}_k$ have at least $r$ neighbors outside of $\mathcal{S}_k$. Given a pair $\mathcal{S}_1,\mathcal{S}_2 \subset \mathcal{V}$ such that $0<|\mathcal{S}_1|<r$ and $\mathcal{S}_2=\mathcal{V}\setminus\mathcal{S}_1$, there can be no more than $|\mathcal{S}_1|$ nodes with at least $r$ neighbors outside of their set. Hence, conditions $(i)$ and $(ii)$ quantify the maximum number of nodes with at least $r$ neighbors outside of their set for such pairs, and must therefore be ``enough''. Alternatively, if there are at least $s$ nodes with at least $r$ neighbors outside of their respective sets in the union $\mathcal{S}_1 \cup \mathcal{S}_2$, then condition $(iii)$ is satisfied. For such pairs $\mathcal{S}_1,\mathcal{S}_2 \subset \mathcal{V}$, the parameter\footnote{Note that $s=0$ is {\it not} allowed in $(r,s)$-robustness because in that case any digraph on $n\geq2$ nodes satisfies the definition for any $r\in\mathbb{Z}_{\geq0}$, which subverts the interpretation of the parameter $r$. At the other extreme, the maximal meaningful value of $s$ is $s=n$ since condition $(iii)$ can {\it never} be satisfied with $s>n$.} $1\leq s\leq n$ quantifies what is meant by ``enough'' nodes.

\begin{figure}[Illustration of Robustness]
\small
\centering
\includegraphics[width=4.2cm]{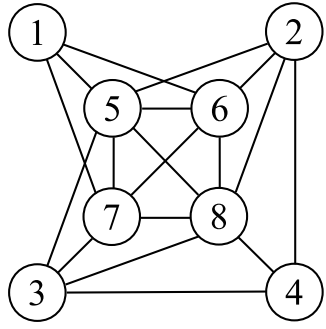}
\caption{A 3-robust graph that is {\it not} (3,2)-robust.}
\label{fig:illustration_robustness}
\end{figure}

An important observation is that $(r,1)$-robustness is equivalent to $r$-robustness. This holds because conditions $(i)-(iii)$ for $(r,1)$-robustness collapse to the condition that at least one of $\mathcal{S}_1$ and $\mathcal{S}_2$ is $r$-reachable. In general, a digraph is $(r,s')$-robust if it is $(r,s)$-robust for $s'\leq s$; therefore, a digraph is $r$-robust whenever it is $(r,s)$-robust. The converse, however, is not true. Consider the graph in Fig.~\ref{fig:illustration_robustness}. This graph is $3$-robust, but is not $(3,2)$-robust. For example, let $\mathcal{S}_1=\{1,3,5,6,7\}$ and $\mathcal{S}_2=\{2,4\}$. Thus, only node 2 has at least 3 nodes outside of its set, so all of the conditions $(i)-(iii)$ fail. Therefore, $(r,s)$-robustness is a strict generalization of $r$-robustness.

Next, consider again the example of Fig.~\ref{fig:Counterexample}. It can be shown that this graph is $(2,s)$-robust, for all $1\leq s \leq n=8$. This follows because {\it all} nodes in at least one of the sets $\mathcal{S}_1$ and $\mathcal{S}_2$ has at least 2 neighbors outside of their set, for any nonempty and disjoint $\mathcal{S}_1,\mathcal{S}_2 \subset \mathcal{V}$. Therefore, condition $(iii)$ in Definition~\ref{D:RTotalRobust} is {\it never} needed, and the definition is satisfied with $r=2$ for all valid values of $s$.

On the other hand, the graph in Fig.~\ref{fig:Counterexample} is {\it not} 3-robust. This can be shown by selecting $\mathcal{S}_1=X$ and $\mathcal{S}_2=Y$. Note that an $(r,s)$-robust digraph is $(r',s)$-robust for $r'\leq r$. The question then arises, how does one compare relative robustness between digraphs? Clearly, if digraph $\mathcal{D}_1$ is ($r_1,s_1$)-robust and digraph $\mathcal{D}_2$ is ($r_2,s_2$)-robust with maximal 
$r_k$ and $s_k$ for $k\in \{1,2\}$, where $r_1>r_2$ and $s_1>s_2$, then one can conclude that $\mathcal{D}_1$ is more robust than $\mathcal{D}_2$. However, in cases where $r_1>r_2$ but $s_1<s_2$, which digraph is more robust? For example, the graph of Fig.~\ref{fig:Counterexample} is ($2,s$)-robust for all $1\leq s\leq n=8$, but is not 3-robust, whereas the graph in Fig.~\ref{fig:illustration_robustness} is 3-robust, but is not (2,5)-robust (e.g., let $\mathcal{S}_1=\{1,5,6\}$ and $\mathcal{S}_2=\{2,3,4\}$). In general, the $r$-robustness property takes precedence in the partial order that determines relative robustness, and the maximal $s$ in ($r,s$)-robustness is used for finer grain partial ordering (i.e., ordering the robustness of two $r$-robust digraphs with the same value of $r$). Therefore, the graph in Fig.~\ref{fig:illustration_robustness} is more robust than the graph of Fig.~\ref{fig:Counterexample}. Yet, the graph of Fig.~\ref{fig:illustration_robustness} is only 3-connected, whereas the graph of Fig.~\ref{fig:Counterexample} is 5-connected. Hence, it is possible that a digraph with {\it less} connectivity is {\it more} robust.

We demonstrate in Section~\ref{S:Results} that the $r$-robustness property is useful for analyzing ARC-P with parameter $F$ under the $F$-local model, and show that $(r,s)$-robustness is the key property for analyzing ARC-P with parameter $F$ under the $F$-total model. More specifically, we show that $(F+1,F+1)$-robustness of the network is both necessary and sufficient for normal nodes using ARC-P with parameter $F$ to achieve resilient asymptotic consensus whenever the scope of threat is $F$-total, the threat model is malicious, and the network is time-invariant. Likewise, we show that $(2F+1)$-robustness of the network is sufficient for ARC-P with parameter $F$ to achieve resilient asymptotic consensus whenever the scope of threat is $F$-local.

\subsection{Construction of Robust Digraphs}
Note that robustness requires checking every possible nonempty disjoint pair of subsets of nodes in the digraph for certain conditions. Currently, we do not have a computationally efficient method to check whether these properties hold in arbitrary digraphs. However, in \cite{Zhang2012_RobInfoDiffAlgs2LocBdAdv} it is shown that the common {\it preferential-attachment} model for complex networks (e.g., \cite{Albert02_StatMechCompNets}) produces $r$-robust graphs, provided that a sufficient number of links are added to the network as new nodes are attached.  In this subsection, we extend this construction to show that preferential-attachment also leads to $(r,s)$-robust graphs.

\begin{theorem}
Let $\mathcal{D}=\{\mathcal{V},\mathcal{E}\}$ be a nonempty, nontrivial ($r,s$)-robust digraph. Then the digraph $\mathcal{D}'=\{\mathcal{V} \cup \{v_{new}\},\mathcal{E} \cup \mathcal{E}_{new}\}$, where $v_{new}$ is a new vertex added to $\mathcal{D}$ and $\mathcal{E}_{new}$ is the directed edge set related to $v_{new}$, is ($r,s$)-robust if $d_{v_{new}}\ge r+s-1$. 
\label{T:GrowRobust}
\end{theorem}
\begin{IEEEproof}
For any pair of nonempty, disjoint sets $\mathcal{S}_1$ and $\mathcal{S}_2$, there are three cases to check: $v_{\text{new}} \not\in \mathcal{S}_i$ , $\{v_{\text{new}}\}=\mathcal{S}_i$ and $v_{\text{new}} \in \mathcal{S}_i$, $i\in \{1,2\}$. In the first case, since $\mathcal{D}$ is $(r,s)$-robust, the conditions in Definition~\ref{D:RTotalRobust} must hold. In the second case, $\mathcal{X}_{\mathcal{S}_i}=\mathcal{S}_i$, and we are done. In the third case, suppose, without loss of generality, $\mathcal{S}_2=\mathcal{S}_2' \cup \{v_{\text{new}} \}$. Since $\mathcal{D}$ is $(r,s)$-robust, at least one of the following conditions hold: $s_{r,1}+s'_{r,2}\ge s$, $s_{r,1}=|\mathcal{S}_1|$, or $s'_{r,2}=|\mathcal{S}_2'|$. If either of the first two hold, then the corresponding conditions hold for the pair $\mathcal{S}_1, \mathcal{S}_2$ in $\mathcal{D'}$. So assume only $s'_{r,2}=|\mathcal{S}_2'|$ holds. Then, the negation of the first condition $s_{r,1}+s'_{r,2}\ge s$ implies $s'_{r,2}=|\mathcal{S}_2'|< s$. Hence, $|\mathcal{V}_{v_{\text{new}}} \setminus \mathcal{S}_2|\geq r$, and $s_{r,2}=|\mathcal{S}_2|$, completing the proof. 
\end{IEEEproof}

The above result indicates that to construct an ($r,s$)-robust digraph with $n$ nodes (where $n > r$), we can start with an ($r,s$)-robust digraph with relatively smaller order (such as a complete graph), and continually add new nodes with incoming edges from at least $r+s-1$ nodes in the existing digraph.  Note that this method does not specify {\it which} existing nodes should be chosen.  The preferential-attachment model corresponds to the case when the nodes are selected with a probability proportional to the number of edges that they already have. This leads to the formation of so-called {\it scale-free} networks \cite{Albert02_StatMechCompNets}, and is cited as a plausible mechanism for the formation of many real-world complex networks.  Theorem~\ref{T:GrowRobust} indicates that a large class of scale-free networks are resilient to the threat models studied in this paper (provided the number of edges added in each round is sufficiently large when the network is forming).

\begin{figure}
\small
\centering
\includegraphics[width=7.4cm]{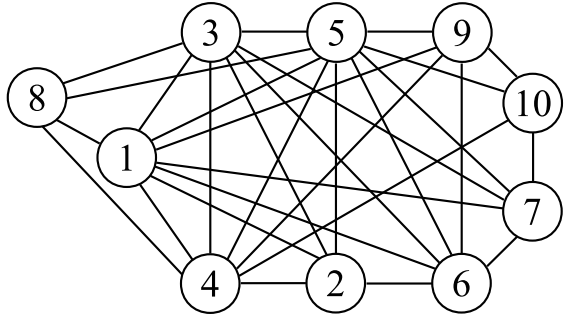}
\caption{A $(3,2)$-robust graph constructed from $K_5$ using preferential attachment.}
\label{fig:ConstructRobustGraph}
\end{figure}

For example, Fig.~\ref{fig:ConstructRobustGraph} illustrates a ($3,2$)-robust graph constructed using the preferential attachment model starting with the complete graph on 5 nodes, $K_5$ (which is also (3,3)-robust and is the only (3,2)-robust digraph on 5 nodes), and with 4 new edges added to each new node. Note that this graph is also 4-robust, which could {\it not} be predicted from Theorem~\ref{T:GrowRobust} since $K_5$ is not 4-robust. Therefore, it is actually possible (but not guaranteed) to end up with a {\it more} robust digraph than the initial one using the preferential-attachment growth model.

\section{\uppercase{Resilient Consensus Results}}
\label{S:Results}

In this section, we provide the key results showing that sufficiently robust digraphs guarantee resilient consensus.  We begin with the following result showing that ARC-P always satisfies the safety condition for resilient asymptotic consensus. Recall that $M[t]$ and $m[t]$ are the maximum and minimum values of the {\it normal} nodes at time $t$, respectively.

\begin{lemma}
\label{L:SafeUpdate}
Suppose each normal node updates its value according to ARC-P with parameter $F$ under the $F$-total or $F$-local malicious model. Then, for each normal node $i\in\mathcal{N}$, ${x}_i[t]\in[m[0],M[0]]$ for all $t$, 
regardless of the network topology.
\end{lemma}
\begin{IEEEproof}
The proof for discrete time is straightforward and follows directly from the definitions and the fact that the values in $\mathcal{J}_i[t]\setminus \mathcal{R}_i[t]$ used in the ARC-P update rule lie in the interval $[m[t],M[t]]$ and the update rule in (\ref{eqn:ft_update}) is a convex combination of these values. For continuous time, we have proved this in Lemma~\ref{L:HypercubeInv}.
\end{IEEEproof}
An immediate consequence of Lemma~\ref{L:SafeUpdate} is that $M[\cdot]$ is nonincreasing with time, and $m[\cdot]$ is nondecreasing with time. From this, it follows that the Lyapunov candidate $\Psi[t]=M[t]-m[t]$ is nonincreasing with time. In the following sections, we show that this Lyapunov function decreases over sufficiently large time intervals whenever the normal nodes update their values according to ARC-P, provided the network is sufficiently robust.
 
\subsection{$F$-Total Model}
\begin{theorem}
\label{T:SufficiencyFTotal}
Consider a time-invariant network modeled by a directed graph $\mathcal{D} = \{\mathcal{V},\mathcal{E}\}$ where each normal node updates its value according to ARC-P with parameter $F$.  Then, resilient asymptotic consensus is achieved under the $F$-total malicious model if and only if the network topology is $(F+1,F+1)$-robust.
\end{theorem}
\begin{IEEEproof}
{\it (Necessity)} If $\mathcal{D}$ is not $(F+1,F+1)$-robust, then there are nonempty, disjoint $\mathcal{S}_1, \mathcal{S}_2 \subset \mathcal{V}$ such that none of the conditions $(i)-(iii)$ hold. Suppose the initial value of each node in $\mathcal{S}_1$ is $a$ and each node in $\mathcal{S}_2$ is $b$, with $a<b$. Let all other nodes have initial values taken from the interval $(a,b)$. Since $s_{F+1,1} + s_{F+1,2} \leq F$, suppose all nodes in $\mathcal{X}_{\mathcal{S}_1}$ and $\mathcal{X}_{\mathcal{S}_2}$ are malicious and keep their values constant. With this assignment of adversaries, there is still at least one normal node in both $\mathcal{S}_1$ and $\mathcal{S}_2$ since $s_{F+1,1} < |\mathcal{S}_1|$ and $s_{F+1,2} < |\mathcal{S}_2|$, respectively. Since these normal nodes remove the $F$ or less values of in-neighbors outside of their respective sets, no consensus among normal nodes is reached.

{\it (Sufficiency)} \textbf{[Continuous Time]}
We know from Lemma~\ref{L:SafeUpdate} that both $M[\cdot]$ and $m[\cdot]$ are monotone and bounded functions of $t$, and therefore each of them has a limit, denoted by $A_M$ and $A_m$, respectively.   Note that if $A_M=A_m$, then the normal nodes will achieve resilient asymptotic consensus.  We will prove by contradiction that this must be the case. The main idea behind the proof is to use the gap between $A_M$ and $A_m$ and combine this with both the uniform continuity assumption on the malicious nodes' value trajectories and a careful selection of subsets of nodes to show that $\Psi[t]$ will shrink to be smaller than the gap $A_M-A_m$ in finite time (a contradiction).  To this end, suppose that $A_M \ne A_m$ (note that $A_M > A_m$ by definition). Since $M[t]\rightarrow A_M$ monotonically, we have $M[t]\geq A_M$ for all $t\geq 0$. Similarly, $m[t]\leq A_m$ for all $t\geq0$. Moreover, for each $\epsilon>0$ there exists $t_\epsilon>0$ such that $M[t] <  A_M + \epsilon$ and $m[t] >  A_m - \epsilon$, $\forall t\ge t_{\epsilon}$.  Next, define constant $\epsilon_0 = (A_M-A_m)/4> 0$, which satisfies $M[t] - \epsilon_0 \geq m[t] + \epsilon_0 + (A_M-A_m)/2$. This inequality informs the choice of subsets of nodes to be defined shortly in order to limit the influence of the malicious nodes. Indeed, since the adversary trajectory $x_k$ is uniformly continuous on $[0,\infty)$ for $k\in\mathcal{A}$, it follows that for each $\nu>0$, there exists $\delta_k(\nu)>0$ such that $|x_k[t_1]-x_k[t_2]|<\nu$ whenever $|t_1-t_2|<\delta_k(\nu)$. Define $\delta(\nu)=\min_{k\in\mathcal{A}}\{\delta_k(\nu)\}$. 

Next, we define the sets of nodes that are vital to the proof. For any $t_0\geq0$, $t\geq t_0$, $\Delta> 0$, and $\eta>0$, define 
\begin{equation*}
\mathcal{X}_{M}(t,t_0,\Delta,\eta)\hspace{-0.1cm}=\hspace{-0.1cm}\{i\in\mathcal{V}\colon \exists t' \in [t,t+\Delta] \text{ s.t. } x_i[t']>M[t_0]-\eta \}
\end{equation*}
and 
\begin{equation*}
\mathcal{X}_{m}(t,t_0,\Delta,\eta)\hspace{-0.1cm}=\hspace{-0.1cm}\{i\in\mathcal{V}\colon \exists t' \in [t,t+\Delta] \text{ s.t. } x_i[t']<m[t_0]+\eta \}.
\end{equation*}
Observe that if we choose $\eta\leq\epsilon_0=(A_M-A_m)/4$, $\nu<(A_M-A_m)/2$, and $\Delta<\delta(\nu)$, then we are guaranteed that for any $t_0\geq0$ and $t\geq t_0$, $\mathcal{X}_{M}(t,t_0,\Delta,\eta)\cap\mathcal{X}_{m}(t,t_0,\Delta,\eta)\cap \mathcal{A}=\emptyset$. That is, with these choices of $\eta$, $\nu$, and $\Delta$, no malicious node can be in both $\mathcal{X}_{M}(t,t_0,\Delta,\eta)$ and $\mathcal{X}_{m}(t,t_0,\Delta,\eta)$. This follows because otherwise there exists $t_1,t_2\in[t,t+\Delta]$ and $k\in\mathcal{A}$ such that $x_k[t_1]>M[t_0]-\eta$ and $x_k[t_2]<m[t_0]+\eta$, from which we reach the contradiction
\begin{equation*}
x_k[t_1]-x_k[t_2]>M[t_0]-m[t_0]-2\eta \geq \frac{A_M-A_m}{2} > \nu.
\end{equation*}

We proceed by showing that if we choose $\eta$, $\nu$, and $\Delta$ small enough, 
then no normal node can be in both $\mathcal{X}_{M}(t,t_0,\Delta,\eta)$ and $\mathcal{X}_{m}(t,t_0,\Delta,\eta)$ for any $t_0\geq0$ and $t\geq t_0$. First, we require some generic bounds on the normal node trajectories. For $i\in\mathcal{N}$ with $x_i[t']<M[t']$, we know from Lemmas \ref{L:BoundAveragePhi} and \ref{L:SafeUpdate} that for $t\geq t'$,
\begin{equation*}
\dot{x}_i[t]= \hspace{-0.2cm} \sum_{j\in \mathcal{V}_i\setminus\mathcal{R}_i[t]} \hspace{-0.1cm}w_{ij}[t]\left(x_j[t]-x_i[t]\right) \leq B(M[t']-x_i[t]),
\end{equation*}
whenever the derivative exists\footnote{The solutions of the normal nodes' trajectories are understood in the sense of Carath\'{e}odory. Hence, it is possible that the derivative of the solution does not exist on a set of points in time of Lebesgue measure zero.}, where $B =(n-F-1)\beta$ is the product of the upper bound on the weights $\beta$ and the maximum number of neighboring values used that have value $M[t]\leq M[t_0]$, $n-1-F$ (since there is a maximum of $n-1$ neighbors, $F$ of which would be thrown away). Using the integrating factor $e^{B(t-t')}$, and integrating in the sense of Lebesgue, we have
\begin{equation}
\label{E:NormalFTIncBd}
x_i[t] \leq x_i[t']e^{-B(t-t')}+M[t'](1-e^{-B(t-t')}), \ \forall t\geq t'.
\end{equation}
By interchanging the roles of $t$ and $t'$, we have
\begin{equation}
\label{E:NormalBTDecBd}
x_i[t] \geq x_i[t']e^{B(t'-t)}+M[t](1-e^{B(t'-t)}), \ \forall t\leq t'.
\end{equation}
Similarly, we can show that for $j\in\mathcal{N}$ with $x_j[t']>m[t']$,
\begin{equation}
\label{E:NormalFTDecBd}
x_j[t] \geq x_j[t']e^{-B(t-t')}+m[t'](1-e^{-B(t-t')}), \ \forall t\geq t',
\end{equation}
and
\begin{equation}
\label{E:NormalBTIncBd}
x_j[t] \leq x_j[t']e^{B(t'-t)}+m[t](1-e^{B(t'-t)}), \ \forall t\leq t',
\end{equation}
Now fix $\eta\leq\epsilon_0=(A_M-A_m)/4$, $\nu<(A_M-A_m)/2$, and $\Delta<\min\{\delta(\nu),\log(3)/B\}$, and suppose $i\in\mathcal{N}\cap\mathcal{X}_{M}(t,t_0,\Delta,\eta)$. Then $\exists t'\in[t,t+\Delta]$ such that $x_i[t']>M[t_0]-\eta$. Combining this with (\ref{E:NormalFTDecBd}), it follows that for $s\in[t',t+\Delta]$,
\begin{align}\nonumber
x_i[s]&\geq x_i[t']e^{-B(s-t')}+m[t'](1-e^{-B(s-t')})\\ \nonumber
&> (M[t_0]-\eta)e^{-B(s-t')}+m[t_0](1-e^{-B(s-t')})\\ \nonumber
&\geq (A_M-\eta)e^{-B(s-t')}+m[t_0]-A_m e^{-B(s-t')}\\ \nonumber
&\geq m[t_0]+(A_M-A_m)e^{-B(s-t')}-\frac{A_M-A_m}{4} e^{-B(s-t')}\\ \nonumber
&\geq m[t_0]+\frac{3}{4}(A_M-A_m)e^{-B\Delta}\\ \nonumber
&> m[t_0]+\frac{A_M-A_m}{4}\geq m[t_0]+\eta, \nonumber
\end{align}
where we have used the fact that $\Delta<\log(3)/B$ in deriving the last line. Similarly, using (\ref{E:NormalBTDecBd}), it follows that for $s\in[t,t']$,
\begin{align}\nonumber
x_i[s]&\geq x_i[t']e^{B(t'-s)}+M[s](1-e^{B(t'-s)})\\ \nonumber
&> (M[t_0]-\eta)e^{B(t'-s)}+M[s](1-e^{B(t'-s)})\\ \nonumber
&\geq M[s]-\eta e^{B(t'-s)}\\ \nonumber
&\geq M[s]-\eta\\ \nonumber
&\geq A_M-\frac{A_M-A_m}{4}\\ \nonumber
&= \frac{A_M+A_m}{2}+\frac{A_M-A_m}{4}\\ \nonumber
&\geq m[t_0]+\eta. \nonumber
\end{align}
Therefore, $i\notin\mathcal{X}_{m}(t,t_0,\Delta,\eta)$. 

Similarly, with the given choices for $\eta$, $\nu$, and $\Delta$, if $j\in\mathcal{N}\cap\mathcal{X}_{m}(t,t_0,\Delta,\eta)$, then  it follows from (\ref{E:NormalFTIncBd}) that for $s\in[t',t+\Delta]$, 
\begin{align}\nonumber
x_j[s]&\leq x_j[t']e^{-B(s-t')}+M[t'](1-e^{-B(s-t')})\\ \nonumber
&< (m[t_0]+\eta)e^{-B(s-t')}+M[t_0](1-e^{-B(s-t')})\\ \nonumber
&\leq M[t_0]-(M[t_0]-m[t_0])e^{-B(s-t')}+\eta e^{-B(s-t')}\\ \nonumber
&\leq M[t_0]-(A_M-A_m)e^{-B(s-t')}+\frac{A_M-A_m}{4} e^{-B(s-t')}\\ \nonumber
&\leq M[t_0]-\frac{3}{4}(A_M-A_m)e^{-B\Delta}\\ \nonumber
&< M[t_0]-\frac{A_M-A_m}{4}\leq M[t_0]-\eta, \nonumber
\end{align}
where we have used the fact that $\Delta<\log(3)/B$ in deriving the last line. Finally, using (\ref{E:NormalBTIncBd}), it follows that for $s\in[t,t']$,
\begin{align}\nonumber
x_j[s]&\leq x_j[t']e^{B(t'-s)}+m[s](1-e^{B(t'-s)})\\ \nonumber
&< (m[t_0]+\eta)e^{B(t'-s)}+m[s](1-e^{B(t'-s)})\\ \nonumber
&\leq m[s]+\eta e^{B(t'-s)}\\ \nonumber
&\leq m[s]+\eta\\ \nonumber
&\leq A_m+\frac{A_M-A_m}{4}\\ \nonumber
&= \frac{A_M+A_m}{2}-\frac{A_M-A_m}{4}\\ \nonumber
&\leq M[t_0]-\eta. \nonumber
\end{align}
Thus, $j\notin\mathcal{X}_{M}(t,t_0,\Delta,\eta)$. This shows that $\mathcal{X}_{M}(t,t_0,\Delta,\eta)$ and $\mathcal{X}_{m}(t,t_0,\Delta,\eta)$ are disjoint for appropriate choices of the parameters.

Next, we show that by choosing $\epsilon$ small enough, we can define a sequence of sets, $\{\mathcal{X}_{M}(t_\epsilon+k\Delta,t_\epsilon,\Delta,\epsilon_k)\}_{k=0}^{k=N}$ and $\{\mathcal{X}_{m}(t_\epsilon+k\Delta,t_\epsilon,\Delta,\epsilon_k)\}_{k=0}^{k=N}$, where $N=|\mathcal{N}|$, so that we are guaranteed that by the $N$th step, at least one of the sets contains no normal nodes. This will be used to show that $\Psi$ has shrunk below $A_M-A_m$. Toward this end, let $\epsilon_0=(A_M-A_m)/4$, $\nu<(A_M-A_m)/2$, and $\Delta<\min\{\delta(\nu),\log(3)/B\}$. Then fix 
\begin{equation*}
\label{E:EpsDefinedCTFTotal}
\epsilon < \frac{1}{2}\left[\frac{\alpha}{B}(1-e^{-B\Delta})e^{-B\Delta}\right]^{2N}\epsilon_0.
\end{equation*}
For $k=0,1,2,\dotsc,N$, define $\epsilon_k=[\tfrac{\alpha}{B}(1-e^{-B\Delta})e^{-B\Delta}]^{2k}\epsilon_0$, which results in $\epsilon_0>\epsilon_1>\dots>\epsilon_N>2\epsilon>0$. For brevity, define $\mathcal{X}_{M}^k=\mathcal{X}_{M}(t_\epsilon+k\Delta,t_\epsilon,\Delta,\epsilon_k)$ and $\mathcal{X}_{m}^k=\mathcal{X}_{m}(t_\epsilon+k\Delta,t_\epsilon,\Delta,\epsilon_k)$ for $k=0,1,\dotsc,N$. Observe that by definition, there is at least one normal node (the ones with extremal values) in $\mathcal{X}_{M}^0$ and $\mathcal{X}_{m}^0$, and we have shown above that all of the $\mathcal{X}_{M}^k$ and $\mathcal{X}_{m}^k$ are disjoint. It follows from the fact that there are at most $F$ malicious nodes in the network ($F$-total model) and $\mathcal{D}$ is $(F+1,F+1)$-robust, that either $\exists i\in \mathcal{X}_M^0\cap\mathcal{N}$ or $\exists i\in \mathcal{X}_m^0\cap\mathcal{N}$ (or both) such that $i$ has at least $F+1$ neighbors outside of its set.  That is, either $i$ has at least $F+1$ neighbors $i_1,i_2,\dotsc, i_{F+1}$ such that $x_{i_k}[t]\leq M[t_\epsilon]-\epsilon_0$ for all $t\in[t_{\epsilon},t_{\epsilon}+\Delta]$ (if $i\in \mathcal{X}_M^0\cap\mathcal{N}$), or $x_{i_k}[t]\geq m[t_\epsilon]+\epsilon_0$ for all $t\in[t_{\epsilon},t_{\epsilon}+\Delta]$ (if $i\in \mathcal{X}_m^0\cap\mathcal{N}$). Note that it can be shown that the minimum in-degree of an $(F+1,F+1)$-robust digraph is at least $2F+1$. It follows from this that $i$ will always use at least one neighbor's value in its update. Assume $i\in \mathcal{X}_M^0\cap\mathcal{N}$ and suppose that none of the $F+1$ (or more) neighbors outside of $\mathcal{X}_M^0$ are used in its update at some time $t'\in[t_\epsilon,t_\epsilon+\Delta]$ at which the derivative exists. Then, $x_i[t']\leq M[t_0]-\epsilon_0$ (otherwise, it would use at least one of its $F+1$ neighbors' values outside of $\mathcal{X}_M^0$. It follows from (\ref{E:NormalFTIncBd}) that
\begin{equation*}
x_i[t_\epsilon+\Delta]\leq  M[t_\epsilon]-\epsilon_0 e^{-B\Delta}.
\end{equation*}
Using this with (\ref{E:NormalFTIncBd}) to upper bound $x_i[t]$ for $t\in[t_\epsilon+\Delta, t_\epsilon+2\Delta]$, we see that
$$
x_i[t]\leq  M[t_\epsilon]-\epsilon_0 e^{-2B\Delta}\leq M[t_\epsilon]-\epsilon_1.
$$
Therefore, in this case $i\notin \mathcal{X}_M^1$.  
Alternatively, assume at least one of the values from its neighbors outside of $\mathcal{X}_M^0$ is used for almost all $t\in[t_\epsilon, t_\epsilon+\Delta]$. Then,
\begin{align}\nonumber
\dot{x}_i[t]&\leq \alpha(M[t_\epsilon]-\epsilon_0-x_i[t])+(B-\alpha)(M[t_\epsilon]-x_i[t])\\ \nonumber
&\leq -B x_i[t]+BM[t_\epsilon]-\alpha \epsilon_0,\nonumber
\end{align}
for almost all $t\in[t_\epsilon, t_\epsilon+\Delta]$. Using this, we can show
\begin{align}\nonumber
x_i[t_\epsilon+\Delta]&\leq x_i[t_\epsilon]e^{-B\Delta}+(M[t_\epsilon]-\tfrac{\alpha \epsilon_0}{B})(1-e^{-B\Delta})\\ \nonumber
&\leq M[t_\epsilon]-\tfrac{\alpha}{B}(1-e^{-B\Delta})\epsilon_0.\nonumber
\end{align}
Using this with (\ref{E:NormalFTIncBd}) to upper bound $x_i[t]$ for $t\in[t_\epsilon+\Delta, t_\epsilon+2\Delta]$, we see that for all $t\in[t_\epsilon+\Delta, t_\epsilon+2\Delta]$,
\begin{align}\nonumber
x_i[t]
&\leq M[t_\epsilon]-\tfrac{\alpha}{B}(1-e^{-B\Delta})e^{-B(t-t_\epsilon-\Delta)}\epsilon_0 \\ \nonumber
&\leq M[t_\epsilon]-\tfrac{\alpha}{B}(1-e^{-B\Delta})e^{-B\Delta}\epsilon_0 \\ \nonumber
&\leq M[t_\epsilon]-\epsilon_1. \\ \nonumber
\end{align}
Thus, in either case $i\notin \mathcal{X}_M^1$. The final step is to show that $j\notin\mathcal{X}_m^1$ whenever $j$ is a normal node with $j\notin\mathcal{X}_m^0$. Since $j\notin\mathcal{X}_m^0$, it means that $x_j[t_\epsilon+\Delta]\geq m[t_\epsilon]+\epsilon_0$. Using this with (\ref{E:NormalFTDecBd}) to lower bound $x_j[t]$ for $t\in[t_\epsilon+\Delta, t_\epsilon+2\Delta]$, we see that
$$
x_j[t]\geq m[t_\epsilon]+\epsilon_0 e^{-B\Delta} \geq m[t_\epsilon]+\epsilon_1.
$$
Hence, $j$ is also not in $\mathcal{X}_m^1$, as claimed.  Therefore, if $i\in \mathcal{X}_M^0\cap\mathcal{N}$ has at least $F+1$ neighbors outside of its set, we are guaranteed that $|\mathcal{X}_M^1\cap \mathcal{N}|<|\mathcal{X}_M^0\cap \mathcal{N}|$ and $|\mathcal{X}_m^1\cap \mathcal{N}|\leq|\mathcal{X}_m^0\cap \mathcal{N}|$. Using a similar argument, we can show that if $i\in \mathcal{X}_m^0\cap\mathcal{N}$ has at least $F+1$ neighbors outside of its set, we are guaranteed that $|\mathcal{X}_m^1\cap \mathcal{N}|<|\mathcal{X}_m^0\cap \mathcal{N}|$ and $|\mathcal{X}_M^1\cap \mathcal{N}|\leq|\mathcal{X}_M^0\cap \mathcal{N}|$. 

Now, if both $\mathcal{X}_M^1\cap\mathcal{N}$ and $\mathcal{X}_m^1\cap\mathcal{N}$ are nonempty, we can repeat the above argument to show that either $|\mathcal{X}_m^2\cap \mathcal{N}|<|\mathcal{X}_m^1\cap \mathcal{N}|$ or $|\mathcal{X}_M^2\cap \mathcal{N}|<|\mathcal{X}_M^1\cap \mathcal{N}|$, or both. It follows by induction that as long as both $\mathcal{X}_M^j\cap\mathcal{N}$ and $\mathcal{X}_m^j\cap\mathcal{N}$ are nonempty, then either $|\mathcal{X}_m^{j+1}\cap \mathcal{N}|<|\mathcal{X}_m^j\cap \mathcal{N}|$ or $|\mathcal{X}_M^{j+1}\cap \mathcal{N}|<|\mathcal{X}_M^j\cap \mathcal{N}|$ (or both), for $j=1,2,\dots$. Since $|\mathcal{X}_m^0\cap\mathcal{N}|+|\mathcal{X}_M^0\cap \mathcal{N}|\leq N$, there exists $T<N$ such that at least one of $\mathcal{X}_M^T\cap \mathcal{N}$ and $\mathcal{X}_m^T\cap \mathcal{N}$ is empty. If $\mathcal{X}_M^T\cap \mathcal{N}=\emptyset$, then $M[t_\epsilon+T\Delta]\leq M[t_\epsilon]-\epsilon_T<M[t_\epsilon]-2\epsilon$. Similarly, if $\mathcal{X}_m^T\cap \mathcal{N}=\emptyset$, then $m[t_\epsilon+T\Delta]\geq m[t_\epsilon]+\epsilon_T>m[t_\epsilon]+2\epsilon$. In either case, $\Psi[t_\epsilon+T\Delta]<A_M-A_m$ and we reach the desired contradiction.

{\it (Sufficiency)} \textbf{[Discrete Time]} 
Because $\Psi$ is a nonincreasing function of $t$, whenever the normal nodes are in agreement at time $t_0$, then consensus is maintained for $t\geq t_0$. Therefore, fix $t_0\geq0$ and assume $\Psi[t_0]>0$. 
For $t\geq t_0$ and $\eta>0$, define $\mathcal{X}_M(t,t_0,\eta)=\{j\in\mathcal{V}\colon x_j[t]> M[t_0]-\eta\}$ and $\mathcal{X}_m(t,t_0,\eta)=\{j\in\mathcal{V}\colon x_j[t]< m[t_0]+\eta\}$. Define $\epsilon_0=\Psi[t_0]/2$ and define $\epsilon_j=\alpha \epsilon_{j-1}$ for $j=1,2,\dotsc,N-1$, where $N=\mathcal{N}$. It follows that $\epsilon_j=\alpha^j\epsilon_0>0$. By definition, the sets $\mathcal{X}_m(t_0,t_0,\epsilon_0)$ and $\mathcal{X}_m(t_0,t_0,\epsilon_0)$ are nonempty and disjoint. Because $\mathcal{D}$ is $(F+1,F+1)$-robust and there are at most $F$ malicious nodes in the network ($F$-total model), it follows that either there exists $i\in\mathcal{X}_M(t_0,t_0,\epsilon_0)\cap\mathcal{N}$ or there exists $i\in\mathcal{X}_m(t_0,t_0,\epsilon_0)\cap\mathcal{N}$, or there exists such $i$ in both, such that $i$ has at least $F+1$ neighbors outside of its set.  Therefore, if $i\in\mathcal{X}_M(t_0,t_0,\epsilon_0)\cap\mathcal{N}$ (with at least $F+1$ neighbors outside its set), then
\begin{align}\nonumber
x_i[t_0+1]&=x_i[t_0]+\sum_{j\in\mathcal{J}_i\setminus \mathcal{R}_i[t_0]}w_{ij}[t_0]x_j[t_0]\\ \nonumber 
&\leq \alpha(M[t_0]-\epsilon_0)+(1-\alpha)M[t_0]\\ \nonumber 
&\leq M[t_0]-\alpha\epsilon_0=M[t_0]-\epsilon_1. \nonumber 
\end{align}
Note that for any normal node not in $\mathcal{X}_M(t_0,t_0,\epsilon_0)$, the above inequality holds because any normal node always uses its own value in the update. From this, we conclude $|\mathcal{X}_M(t_0+1,t_0,\epsilon_1)\cap\mathcal{N}|<|\mathcal{X}_M(t_0,t_0,\epsilon_0)\cap\mathcal{N}|$. Similarly, if $i\in\mathcal{X}_m(t_0,t_0,\epsilon_0)\cap\mathcal{N}$ (with at least $F+1$ neighbors outside its set), then
\begin{align}\nonumber
x_i[t_0+1]&=x_i[t_0]+\sum_{j\in\mathcal{J}_i\setminus \mathcal{R}_i[t_0]}w_{ij}[t_0]x_j[t_0]\\ \nonumber 
&\geq \alpha(m[t_0]+\epsilon_0)+(1-\alpha)m[t_0]\\ \nonumber 
&\geq m[t_0]+\alpha\epsilon_0=m[t_0]+\epsilon_1. \nonumber 
\end{align}
Similarly as above, this inequality holds for any normal node not in $\mathcal{X}_m(t_0,t_0,\epsilon_0)$. From this, we conclude
$$
|\mathcal{X}_m(t_0+1,t_0,\epsilon_1)\cap\mathcal{N}|<|\mathcal{X}_m(t_0,t_0,\epsilon_0)\cap\mathcal{N}|.
$$ 

By repeating this analysis, we can show by induction that as long as both $\mathcal{X}_M(t_0+j,t_0,\epsilon_j)\cap\mathcal{N}$ and $\mathcal{X}_m(t_0+j,t_0,\epsilon_j)\cap\mathcal{N}$ are both nonempty, then either $|\mathcal{X}_M(t_0+j+1,t_0,\epsilon_{j+1})\cap\mathcal{N}|<|\mathcal{X}_M(t_0+j,t_0,\epsilon_j)\cap\mathcal{N}|$, or $|\mathcal{X}_m(t_0+j+1,t_0,\epsilon_{j+1})\cap\mathcal{N}|<|\mathcal{X}_m(t_0+j,t_0,\epsilon_j)\cap\mathcal{N}|$, or both. Since $|\mathcal{X}_M(t_0,t_0,\epsilon_0)\cap\mathcal{N}|+|\mathcal{X}_m(t_0,t_0,\epsilon_0)\cap\mathcal{N}|\leq|\mathcal{N}|=N$, there exists $T<N$ such that one of the sets $\mathcal{X}_M(t_0+T,t_0,\epsilon_T)\cap\mathcal{N}$, $\mathcal{X}_m(t_0+T,t_0,\epsilon_T)\cap\mathcal{N}$, or both, is empty. It follows that in the former case, $M[t_0+T]\leq M[t_0]-\epsilon_T$, and in the latter case, $m[t_0+T]\geq m[t_0]+\epsilon_T$. Since $\epsilon_0>\epsilon_1>\dots>\epsilon_T\geq\epsilon_{N-1}>0$, we have
\begin{align}\nonumber
\Psi[t_0+&N-1]-\Psi[t_0]\leq\Psi[t_0+T]-\Psi[t_0]\\ \nonumber 
&\leq (M[t_0+T]-M[t_0])+(m[t_0]-m[t_0+T])\\ \nonumber 
&\leq -\epsilon_T \leq -\epsilon_{N-1}. \nonumber 
\end{align}
Therefore, $\Psi[t_0+N-1]\leq\Psi[t_0](1-\alpha^{N-1}/2)$. Define $c=(1-\alpha^{N-1}/2)$. Since $c$ is not a function of $t_0$ and $t_0$ was chosen arbitrarily, it follows that
$$
\Psi[t_0+k(N-1)]\leq c^k \Psi[t_0],
$$
for all $k\in\mathbb{Z}_0$. Because $c<1$, it follows that $\Psi[t]\rightarrow 0$ as $t\rightarrow \infty$. 
\end{IEEEproof}

When the network is time-varying, one can state the following corollary of the above theorem.

\begin{corollary}
\label{C:FTotalTimeVar}
Consider a time-varying network modeled by a directed graph $\mathcal{D}[t] = \{\mathcal{V},\mathcal{E}[t]\}$ where each normal node updates its value according to ARC-P with parameter $F$.  Then, resilient asymptotic consensus is achieved under the $F$-total malicious model if there exists $t_0 \geq 0$ such that $\mathcal{D}[t]$ is $(F+1,F+1)$-robust, $\forall t\geq t_0$.
\end{corollary}
\begin{IEEEproof}
\textbf{[Continuous Time]}
The proof follows the contradiction argument of the proof of Theorem~\ref{T:SufficiencyFTotal}, but here we use the dwell time assumption. In this case, let 
$$\Delta<\min\{\delta(\nu),\log(3)/B, \tfrac{\tau}{N} \}.$$
Fix 
\begin{equation*}
\epsilon < \frac{1}{2}\left[\frac{\alpha}{B}(1-e^{-B\Delta})e^{-B\Delta}\right]^{2N}\epsilon_0,
\end{equation*}
and let $t_{\epsilon}'\geq 0$ be the time such that $M[t]<A_M+\epsilon$ and $m[t]>A_m-\epsilon$ for all $t\geq t_{\epsilon}'$ and define $t'=\max\{t_0,t_{\epsilon}'\}$. Then, associated to the switching signal $\sigma(t)$, we define $t_{\epsilon}$ as the next switching instance after $t'$, or $t'$ itself if there are no switching instances after $t'$. Since $\Delta<\tau/N$, the same sequence of calculations can be used (as in the proof of Theorem~\ref{T:SufficiencyFTotal}) to show that $\Psi[t_{\epsilon}+T\Delta]<A_M-A_m$. 

\textbf{[Discrete Time]}
The argument in the proof of Theorem~\ref{T:SufficiencyFTotal} holds for $t\geq t_0$. Hence,
$$
\Psi[t_0+k(N-1)]\leq c^k \Psi[t_0],
$$
for all $k\in\mathbb{Z}_0$. Because $c<1$, it follows that $\Psi[t]\rightarrow 0$ as $t\rightarrow \infty$. 
\end{IEEEproof}


To illustrate these results on the examples of Section~\ref{S:Topologies}, the graphs in Figs.~\ref{fig:Counterexample}, \ref{fig:illustration_robustness}, and \ref{fig:ConstructRobustGraph} can withstand the compromise of at most 1 malicious node in the network using ARC-P with parameter $F=1$ (each graph is (2,2)-robust but not (3,3)-robust). This is not to say that it is impossible for the normal nodes to reach consensus if there are, for example, two nodes that are compromised. Instead, these results say that it is not possible that {\it any} two nodes can be compromised and still guarantee resilient asymptotic consensus using ARC-P with parameter $F=2$.

\subsection{$F$-Local Model}
\begin{theorem}
Consider a time-invariant network modeled by a directed graph $\mathcal{D} = \{\mathcal{V},\mathcal{E}\}$ where each normal node updates its value according to ARC-P with parameter $F$.  Then, resilient asymptotic consensus is achieved under the $F$-local malicious model if the network topology is $(2F+1)$-robust. Furthermore, a necessary condition is for the topology of the network to be $(F+1)$-robust.
\label{thm:F_local_sufficient}
\end{theorem}
\begin{IEEEproof}
The necessity proof is given in \cite{Zhang2012_RobInfoDiffAlgs2LocBdAdv}. The sufficiency proof follows the same line as that of Theorem~\ref{T:SufficiencyFTotal}. In continuous time, the main difference is that the sets of nodes $\mathcal{X}_{M}$ and $\mathcal{X}_{m}$ include only normal nodes.  That is, for any $t_0\geq0$, $t\geq t_0$, $\Delta> 0$, and $\eta>0$, define 
\begin{equation*}
\mathcal{X}_{M}(t,t_0,\Delta,\eta)\hspace{-0.1cm}=\hspace{-0.1cm}\{i\in\mathcal{N}\colon \exists t' \in [t,t+\Delta] \text{ s.t. } x_i[t']>M[t_0]-\eta \}
\end{equation*}
and 
\begin{equation*}
\mathcal{X}_{m}(t,t_0,\Delta,\eta)\hspace{-0.1cm}=\hspace{-0.1cm}\{i\in\mathcal{N}\colon \exists t' \in [t,t+\Delta] \text{ s.t. } x_i[t']<m[t_0]+\eta \}.
\end{equation*}
Likewise, for $k=1,2,\dotsc,N$, the definitions of $\mathcal{X}_{M}^k$ and $\mathcal{X}_{m}^k$ are modified to include only normal nodes. The analysis showing that $\mathcal{X}_{M}^k$ and $\mathcal{X}_{m}^k$ are disjoint still holds. By definition, it follows that $\mathcal{X}_{M}^0$ and $\mathcal{X}_{m}^0$ are nonempty. Since the network is $(2F+1)$-robust, either $\exists i\in\mathcal{X}_M^0$ or $\exists i\in\mathcal{X}_m^0$, or both, such that $i$ has at least $2F+1$ neighbors outside of its set. If such $i$ is in $\mathcal{X}_M^0$, then at most $F$ of the neighbors are malicious ($F$-local model) and the others are normal with value $x_j[t]\leq M[t_{\epsilon}]-\epsilon_0$ for $t\in[t_\epsilon,t_\epsilon+\Delta]$. The remaining argument follows the same line as that of Theorem~\ref{T:SufficiencyFTotal}.  (Notice in this case that the uniform continuity assumption on the malicious nodes is not needed).

In discrete time, the sets $\mathcal{X}_{M}$ and $\mathcal{X}_{m}$ are defined to include only normal nodes. Then, the $(2F+1)$-robust assumption under the $F$-local model ensures at least one normal value outside of either $\mathcal{X}_{M}$ or $\mathcal{X}_{m}$ will be used in the update. The rest of the analysis is identical to the proof of Theorem~\ref{T:SufficiencyFTotal}.
\end{IEEEproof}

As with the $F$-total model, we have the following corollary (whose proof follows the same line as that of Corollary~\ref{C:FTotalTimeVar}).
\begin{corollary}
Consider a time-varying network modeled by a directed graph $\mathcal{D}[t] = \{\mathcal{V},\mathcal{E}[t]\}$ where each normal node updates its value according to ARC-P with parameter $F$.  Then, resilient asymptotic consensus is achieved under the $F$-local malicious model if there exists $t_0 \geq 0$ such that $\mathcal{D}[t]$ is $(2F+1)$-robust, $\forall t\geq t_0$.
\end{corollary}

To illustrate these results, consider the 3-robust graph of Fig.~\ref{fig:illustration_robustness}. Recall that this graph cannot generally sustain 2 malicious nodes as specified by the 2-total model; it is not (3,3)-robust. However, under the 1-local model, it can sustain two malicious nodes if the {\it right} nodes are compromised. For example, nodes 1 and 4 may be compromised under the 1-local model and the normal nodes will still reach consensus. This example illustrates the advantage of the $F$-local model, where there is no concern about global assumptions. If a digraph is $(2F+1)$-robust, then up to $F$ nodes may be compromised in any node's neighborhood, possibly resulting in more than $F$ malicious nodes in the network (as in the previous example).

\section{\uppercase{Simulations}}
\label{S:Simulations}
\begin{figure}[htb]
\begin{center}
\includegraphics[width=8.2cm]{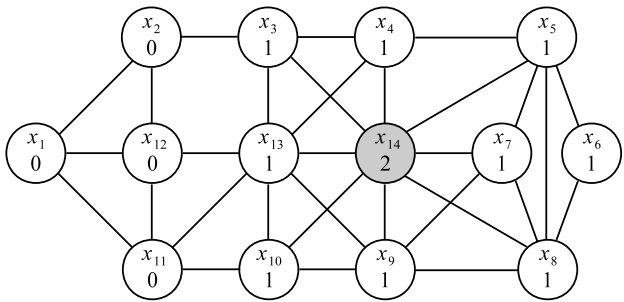}
\caption{\label{fig:MalTotalRobustTop} (2,2)-Robust Network topology.}
\end{center}
\vspace{-0.2cm}
\end{figure}

This section presents a numerical example to illustrate our results. In this example, the network is given by the (2,2)-robust graph shown in Fig.~\ref{fig:MalTotalRobustTop}. Since the network is (2,2)-robust, it can sustain a single malicious node in the network under the 1-total model. Suppose that the node with the largest degree, node 14, is compromised and turns malicious. The nodes have continuous dynamics and the normal nodes use either the Linear Consensus Protocol (LCP) given in (\ref{E:LCP}) or ARC-P for their control input. In either case, the weights are selected to be unity for all neighboring nodes that are kept, with the self-weights selected as $-d_i$ for LCP and $|\mathcal{R}_i[t]|-d_i$ for ARC-P for each normal node $i\in\mathcal{N}$. The initial values of the nodes are shown in Fig.~\ref{fig:MalTotalRobustTop} beneath the label of the node's value. The goal of the malicious agent is to drive the values of the normal nodes to a value of 2.

\begin{figure}[htb]
\centering
	\subfigure[LCP.]{
	  \includegraphics[width=8.2cm]{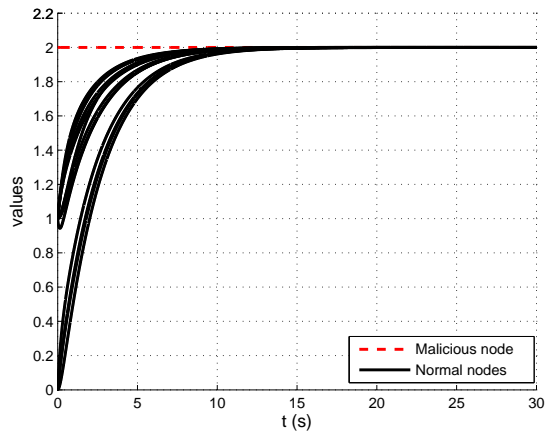}
	  \label{F:MalTotalRobustLCP}}
	\subfigure[ARC-P.]{
	  \includegraphics[width=8.2cm]{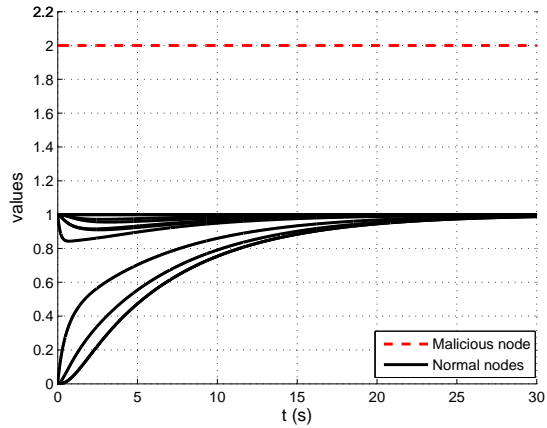}
	   \label{F:MalTotalRobustARCP} }	
\caption{\label{F:MalTotalRobust} Malicious node attempts to drive the values of the normal nodes to a value of 2. The malicious node succeeds whenever LCP is used, but fails whenever ARC-P is used.}
\end{figure}

The results for this example are shown in Fig.~\ref{F:MalTotalRobust}. It is clear in Fig.~\ref{F:MalTotalRobust}(a) that the malicious node is able to drive the values of the normal nodes to its value of 2 whenever LCP is used. On the other hand, the malicious node is unable to achieve its goal whenever ARC-P is used. Note that due to the large degree of the malicious node, it has the potential to drive the consensus process to any value in the interval $[0,1]$ by choosing the desired value as its initial value and remaining constant. However, this is allowed with resilient asymptotic consensus (because the consensus value is within the range of the initial values held by normal nodes). Another observation is that the consensus process in the case of ARC-P is slower than LCP; this is to be expected, due to the fact that ARC-P effectively removes certain edges from the network at each time instance.  

\section{\uppercase{Discussion}}
\label{S:Discussion}
The notion of graph connectivity has long been the backbone of investigations into fault tolerant and secure distributed algorithms. Indeed, under the assumption of full knowledge of the network topology, connectivity is {\it the key} metric in determining whether a fixed number of malicious adversaries can be overcome. However, in large scale systems and complex networks, it is not practical for the various nodes to obtain knowledge of the global network topology. This necessitates the development of algorithms that allow the nodes to operate on purely local information. This paper continues and extends the work started in \cite{LeBlanc11_ConsNetMASWithAdv, LeBlanc_LowCompResConsAdv_HSCC12, Zhang2012_RobInfoDiffAlgs2LocBdAdv}, and represents a step in this direction for the particular application of distributed consensus. Using the ARC-P algorithm developed in \cite{LeBlanc11_ConsNetMASWithAdv}, the notion of robust graphs introduced in \cite{Zhang2012_RobInfoDiffAlgs2LocBdAdv}, and the extensions of each presented here,  we characterize necessary/sufficient conditions for the normal nodes in large-scale networks to mitigate the influence of adversaries. We show that the notions of robust digraphs are the appropriate analogues to graph connectivity when considering purely local filtering rules at each node in the network. Just as the notion of connectivity has played a central role in the existing analysis of reliable distributed algorithms with global topological knowledge, we believe that robust digraphs (and its variants) will play an important role in the investigation of purely local algorithms.

In a recent paper, developed independently of our work, Vaidya {\it et al}.\ have characterized the tight conditions for resilient consensus using only local information whenever the threat model is Byzantine and the scope of threat is $F$-total~\cite{Vaidya2012_ItAppByzConsInArbDigraphs}. The network constructions used in  \cite{Vaidya2012_ItAppByzConsInArbDigraphs} are very similar to the robust digraphs presented here. In particular, the networks in \cite{Vaidya2012_ItAppByzConsInArbDigraphs} also require redundancy of information flow between subsets of nodes in the network in a single hop. 

Finally we summarize the main works related to resilient consensus using only local information in Table~\ref{table:relwork}. In this table, we include only works on resilient consensus (also referred to as Byzantine approximate consensus, or just approximate consensus in the literature) in synchronous networks that use only local information, with no relaying of information across the network and with networks that are {\it not} complete (since complete networks provide global information and have high communication cost). Further discussion about the relationship of the results in this paper (and in \cite{LeBlanc11_ConsNetMASWithAdv, LeBlanc_LowCompResConsAdv_HSCC12, Zhang2012_RobInfoDiffAlgs2LocBdAdv,Vaidya2012_ItAppByzConsInArbDigraphs}) to approximate consensus can be found in \cite{Zhang2012_RobInfoDiffAlgs2LocBdAdv} and \cite{Vaidya2012_ItAppByzConsInArbDigraphs}.

\begin{table}
\centering
\caption{Related work for resilient consensus in synchronous networks using only local information (no nonlocal information, no relays, and the network is {\it not} complete).}
\begin{tabular}{|c|c|c|}
\hline
\backslashbox { Scope }{ Threat Model }            & Byzantine      																																 & Malicious \\
\hline
$F$-total     & \cite{LeBlanc_LowCompResConsAdv_HSCC12, Vaidya2012_ItAppByzConsInArbDigraphs}  & \cite{LeBlanc_LowCompResConsAdv_HSCC12}, this paper\\ \hline
$F$-local     &      --          & \cite{Zhang2012_RobInfoDiffAlgs2LocBdAdv}, this paper \\ \hline
\end{tabular}
\label{table:relwork}
\end{table}

\section{ACKNOWLEDGMENTS}
H. J. LeBlanc and X. Koutsoukos are supported in part by the National Science Foundation (CNS-1035655, CCF-0820088),
the U.S. Army Research Office (ARO W911NF-10-1-0005), and Lockheed Martin. The views and conclusions
contained herein are those of the authors and should not be interpreted as
necessarily representing the official policies or endorsements, either expressed or
implied, of the U.S. Government.  
H. Zhang and S. Sundaram are supported in part by a grant from 
the Natural Sciences and Engineering Research Council of Canada (NSERC), and by a grant from the Waterloo Institute for Complexity and Innovation (WICI).

\end{document}